\documentclass[ALICE,manyauthors]{cernphprep}
\usepackage[comma,square,numbers,sort&compress]{natbib}
\usepackage{hyperref}
\usepackage{lineno}
\usepackage{xspace}
\usepackage[T1]{fontenc}
\usepackage{orcidlink}

\begin{document}
%

\newcommand{\pp}           {pp\xspace}
\newcommand{\ppbar}        {\mbox{$\mathrm {p\overline{p}}$}\xspace}
\newcommand{\XeXe}         {\mbox{Xe--Xe}\xspace}
\newcommand{\PbPb}         {\mbox{Pb--Pb}\xspace}
\newcommand{\pA}           {\mbox{pA}\xspace}
\newcommand{\pPb}          {\mbox{p--Pb}\xspace}
\newcommand{\AuAu}         {\mbox{Au--Au}\xspace}
\newcommand{\dAu}          {\mbox{d--Au}\xspace}

\newcommand{\s}            {\ensuremath{\sqrt{s}}\xspace}
\newcommand{\snn}          {\ensuremath{\sqrt{s_{\mathrm{NN}}}}\xspace}
\newcommand{\pt}           {\ensuremath{p_{\rm T}}\xspace}
\newcommand{\meanpt}       {$\langle p_{\mathrm{T}}\rangle$\xspace}
\newcommand{\ycms}         {\ensuremath{y_{\rm CMS}}\xspace}
\newcommand{\ylab}         {\ensuremath{y_{\rm lab}}\xspace}
\newcommand{\etarange}[1]  {\mbox{$\left | \eta \right |~<~#1$}}
\newcommand{\yrange}[1]    {\mbox{$\left | y \right |~<~#1$}}
\newcommand{\dndy}         {\ensuremath{\mathrm{d}N_\mathrm{ch}/\mathrm{d}y}\xspace}
\newcommand{\dndeta}       {\ensuremath{\mathrm{d}N_\mathrm{ch}/\mathrm{d}\eta}\xspace}
\newcommand{\avdndeta}     {\ensuremath{\langle\dndeta\rangle}\xspace}
\newcommand{\dNdy}         {\ensuremath{\mathrm{d}N_\mathrm{ch}/\mathrm{d}y}\xspace}
\newcommand{\Npart}        {\ensuremath{N_\mathrm{part}}\xspace}
\newcommand{\Ncoll}        {\ensuremath{N_\mathrm{coll}}\xspace}
\newcommand{\dEdx}         {\ensuremath{\textrm{d}E/\textrm{d}x}\xspace}
\newcommand{\RpPb}         {\ensuremath{R_{\rm pPb}}\xspace}

\newcommand{\nineH}        {$\sqrt{s}~=~0.9$~Te\kern-.1emV\xspace}
\newcommand{\seven}        {$\sqrt{s}~=~7$~Te\kern-.1emV\xspace}
\newcommand{\twoH}         {$\sqrt{s}~=~0.2$~Te\kern-.1emV\xspace}
\newcommand{\twosevensix}  {$\sqrt{s}~=~2.76$~Te\kern-.1emV\xspace}
\newcommand{\five}         {$\sqrt{s}~=~5.02$~Te\kern-.1emV\xspace}
\newcommand{\twosevensixnn}{$\sqrt{s_{\mathrm{NN}}}~=~2.76$~Te\kern-.1emV\xspace}
\newcommand{\fivenn}       {$\sqrt{s_{\mathrm{NN}}}~=~5.02$~Te\kern-.1emV\xspace}
\newcommand{\LT}           {L{\'e}vy-Tsallis\xspace}
\newcommand{\GeVc}         {Ge\kern-.1emV/$c$\xspace}
\newcommand{\MeVc}         {Me\kern-.1emV/$c$\xspace}
\newcommand{\TeV}          {Te\kern-.1emV\xspace}
\newcommand{\GeV}          {Ge\kern-.1emV\xspace}
\newcommand{\MeV}          {Me\kern-.1emV\xspace}
\newcommand{\GeVmass}      {Ge\kern-.2emV/$c^2$\xspace}
\newcommand{\MeVmass}      {Me\kern-.2emV/$c^2$\xspace}
\newcommand{\lumi}         {\ensuremath{\mathcal{L}}\xspace}

\newcommand{\ITS}          {\rm{ITS}\xspace}
\newcommand{\TOF}          {\rm{TOF}\xspace}
\newcommand{\ZDC}          {\rm{ZDC}\xspace}
\newcommand{\ZDCs}         {\rm{ZDCs}\xspace}
\newcommand{\ZNA}          {\rm{ZNA}\xspace}
\newcommand{\ZNC}          {\rm{ZNC}\xspace}
\newcommand{\SPD}          {\rm{SPD}\xspace}
\newcommand{\SDD}          {\rm{SDD}\xspace}
\newcommand{\SSD}          {\rm{SSD}\xspace}
\newcommand{\TPC}          {\rm{TPC}\xspace}
\newcommand{\TRD}          {\rm{TRD}\xspace}
\newcommand{\VZERO}        {\rm{V0}\xspace}
\newcommand{\VZEROA}       {\rm{V0A}\xspace}
\newcommand{\VZEROC}       {\rm{V0C}\xspace}
\newcommand{\Vdecay} 	   {\ensuremath{V^{0}}\xspace}

\newcommand{\ee}           {\ensuremath{e^{+}e^{-}}} 
\newcommand{\pip}          {\ensuremath{\pi^{+}}\xspace}
\newcommand{\pim}          {\ensuremath{\pi^{-}}\xspace}
\newcommand{\kap}          {\ensuremath{\rm{K}^{+}}\xspace}
\newcommand{\kam}          {\ensuremath{\rm{K}^{-}}\xspace}
\newcommand{\pbar}         {\ensuremath{\rm\overline{p}}\xspace}
\newcommand{\kzero}        {\ensuremath{{\rm K}^{0}_{\rm{S}}}\xspace}
\newcommand{\lmb}          {\ensuremath{\Lambda}\xspace}
\newcommand{\almb}         {\ensuremath{\overline{\Lambda}}\xspace}
\newcommand{\Om}           {\ensuremath{\Omega^-}\xspace}
\newcommand{\Mo}           {\ensuremath{\overline{\Omega}^+}\xspace}
\newcommand{\X}            {\ensuremath{\Xi^-}\xspace}
\newcommand{\Ix}           {\ensuremath{\overline{\Xi}^+}\xspace}
\newcommand{\Xis}          {\ensuremath{\Xi^{\pm}}\xspace}
\newcommand{\Oms}          {\ensuremath{\Omega^{\pm}}\xspace}
\newcommand{\degree}       {\ensuremath{^{\rm o}}\xspace}

\begin{titlepage}
\PHyear{2023}       
\PHnumber{287}      
\PHdate{11 December}  

\title{Investigating the composition of the K${\bf ^*_0(700)}$ state with $\pi^\pm$K$^0_{\rm S}$ correlations 
at the LHC}
\ShortTitle{$\pi^\pm$K$^0_{\rm S}$ correlations }   
%
\Collaboration{ALICE Collaboration\thanks{See Appendix~\ref{app:collab} for the list of collaboration members}}
\ShortAuthor{ALICE Collaboration} 

\begin{abstract}
The first measurements of femtoscopic correlations with the particle pair combinations $\pi^\pm$K$^0_{\rm S}$
in pp collisions at $\sqrt{s}=13$ TeV at the Large Hadron Collider (LHC) are reported by the 
ALICE experiment. Using the femtoscopic approach, it is shown that it is possible to  study the elusive K$^*_0(700)$ particle that has been considered a tetraquark candidate for over forty years. 
Source and final-state interaction parameters 
are extracted by fitting a model
assuming a Gaussian source to the experimentally measured two-particle correlation functions. The final-state interaction in the $\pi^\pm$K$^0_{\rm S}$ system
 is modeled through a
resonant scattering amplitude, defined in terms of a mass and a coupling parameter. The extracted mass and Breit--Wigner width, derived from the coupling parameter, of the final-state interaction are found to be consistent with previous measurements of the K$^*_0(700)$. The small value and increase of the correlation strength with increasing source size support the hypothesis that the
K$^*_0(700)$ is a four-quark state, i.e. a tetraquark 
state of the form
$({\rm q_1},\overline{\rm q_2}, {\rm q_3}, \overline{\rm q_3})$ in which ${\rm q_1}$, ${\rm q_2}$, and ${\rm q_3}$ indicate the flavor of the valence quarks
of the $\pi$ and K$^0_{\rm S}$.
This latter trend is also confirmed via a simple geometric model that assumes a tetraquark structure of the
K$^*_0(700)$ resonance.

\end{abstract}
\end{titlepage}
\setcounter{page}{2} 


\section{Introduction}

Femtoscopy with identical charged pions has been a useful tool for many years to experimentally 
probe the geometry of
the space-time structure of the freeze-out probability distribution
in high-energy pp and heavy-ion collisions~\cite{Lisa:2005dd}. Identical-kaon femtoscopic measurements
have also been carried out to complement the identical pion studies, examples of which are
measurements in 
Au--Au collisions at center-of-mass energy per nucleon pair $\sqrt{s_{\rm NN}}=200$ GeV at the Relativistic
Heavy-Ion Collider by the STAR Collaboration~\cite{Abelev:2006gu} 
(K$^0_{\rm S}$K$^0_{\rm S}$)
and PHENIX Collaboration~\cite{PHENIX:2015jaj} (K$^\pm$K$^\pm$),
and for pp collisions at $\sqrt{s}=5.02$, $7$, and $13$ TeV and
Pb--Pb collisions at $\sqrt{s_{\rm NN}}=2.76$ TeV at the CERN LHC by the ALICE Collaboration ~\cite{Abelev:2012ms,Abelev:2012sq,Adam:2015vja,ALICE:2021ovd}
(K$^0_{\rm S}$K$^0_{\rm S}$ and K$^{\rm \pm}$K$^{\rm \pm}$).

In the femtoscopic method, the momentum correlations of pairs of particles
when interactions with the other particles in the collision system cease,
i.e. during ``freeze out''~\cite{Fabbietti:2020bfg},
can be utilized to get insight into the strength of the pair interaction,
i.e. the final-state interaction (FSI), at low relative momentum. 
The homogeneity region size, the strength, and even the nature of the FSI at freeze out
can be determined by fitting the experimental two-particle correlation function to a model
based on the FSI.
Results on non-identical kaon femtoscopy with K$^0_{\rm S}$K$^\pm$ pairs were published by ALICE
in pp collisions at $\sqrt{s}=5.02$, $7$, and $13$ TeV and
Pb--Pb collisions at $\sqrt{s_{\rm NN}}=2.76$ TeV~\cite{Acharya:2018kpo,Acharya:2017jks,ALICE:2021ovd}. 
Although the general goals of non-identical kaon femtoscopy studies overlap with those for identical
kaon femtoscopy, e.g. to extract information about the space--time geometry of the collision region and determine
the pair-wise interaction strength,
the latter is different in each case. For the identical kaon cases, in which pair-wise quantum statistical correlations are
present,
the interactions are
the following: K$^{\rm \pm}$K$^{\rm \pm}$ -- Coulomb interaction, and
K$^0_{\rm S}$K$^0_{\rm S}$ -- strong FSI through the f$_0(980)$/a$_0(980)$ resonances.
For the K$^0_{\rm S}$K$^{\rm \pm}$ pairs, there are no quantum statistical correlations and the only interaction present is the strong FSI through the a$_0(980)$ resonance.

K$^0_{\rm S}$K$^{\rm \pm}$ femtoscopy should thus be sensitive to the
properties of the a$_0(980)$ resonance. It has been suggested in many papers
in the literature that the a$_0(980)$ could be a four-quark or tetraquark state\cite{Santopinto:2006my}. 
It was first proposed in 1977 that experimentally-observed low-lying mesons, such as 
the a$_0(980)$ and K$^*_0(700)$, are 
part of a SU(3) tetraquark
nonet using the MIT Bag model~\cite{Jaffe:1976ig}, which was later followed up with lattice QCD calculations~\cite{Alford:2000mm}. 
There have been a number of QCD studies of these mesons that can be categorized as QCD-inspired models, see for example Refs.~\cite{Santopinto:2006my,Narison:2008nj,Achasov:2017zhy,Azizi:2019kzj}, and lattice QCD calculations, see for example 
Refs.~\cite{Dudek:2016cru,Briceno:2016mjc,Guo:2013nja}.

Indeed, the results of the ALICE K$^0_{\rm S}$K$^{\rm \pm}$ studies mentioned above suggested that the a$_0(980)$ is a tetraquark state. This suggestion is based on comparing the extracted pair-wise interaction strength of K$^0_{\rm S}$K$^{\rm \pm}$ between pp and Pb--Pb collisions as well as with the K$^0_{\rm S}$K$^0_{\rm S}$ studies~\cite{Abelev:2012ms,Adam:2015vja,Acharya:2018kpo,Acharya:2017jks,ALICE:2021ovd}.
From a geometric picture, since a tetraquark version of the a$_0(980)$ contains a strange -- anti-strange quark pair,
a FSI through it should be suppressed for a small system as in pp collisions due to an increased annihilation 
probability, whereas for a large Pb--Pb collision this suppression should not be present. Thus, a strong FSI would be
expected from Pb--Pb collisions and weak FSI from pp collisions, and this is what was observed in experiments. It would also be expected that a strong pair-wise correlation would be seen for K$^0_{\rm S}$K$^0_{\rm S}$ studies
since quantum statistics is dominant over FSI effects. This exception is corroborated by experimental findings~\cite{Abelev:2012ms,Adam:2015vja,ALICE:2021ovd}.

The success of the ALICE K$^0_{\rm S}$K$^{\rm \pm}$ studies on the nature of the a$_0(980)$ resonance motivated the first femtoscopic study ever of $\pi^\pm$K$^0_{\rm S}$ correlations in $\sqrt{s}=13$ TeV pp collisions. Another resonance that is a tetraquark candidate is the K$^*_0(700)$ that decays with a branching ratio of $\sim 100\%$ into 
$\pi$K pairs~\cite{Jaffe:1976ig}. 
The K$^*_0(700)$ is listed in the Review of Particle Physics~\cite{Workman:2022ynf} as a strange meson with spin 0 and isospin $\frac{1}{2}$,
the quark content of the K$^*_0(700)^+$ state being u$\overline{\rm s}$. Its mass is listed
as $845\pm17$ MeV/$c^2$ and it is a very broad resonance with Breit--Wigner width of $468\pm30$ MeV/$c^2$.
The mass of the K$^*_0(700)$ is above the $\pi^\pm$K$^0_{\rm S}$ threshold, that is of about 637.18 MeV/$c^2$, and its width is seen to encompass this threshold and below.
The tetraquark version of the
K$^*_0(700)^+$ would have quark content u$\overline{\rm s}$d$\overline{\rm d}$ and would decay by direct
quark transfer into a $\pi^+$K$^0$ pair~\cite{Jaffe:1976ig}.
Thus by measuring $\pi^\pm$K$^0_{\rm S}$ correlations it should be possible to study the quark nature of the  K$^*_0(700)$ using similar methods as mentioned above for the a$_0(980)$ studies, i.e.~measuring the strength
of the FSI, 
assuming that the $\pi^\pm$K$^0_{\rm S}$ FSI goes solely through the K$^*_0(700)$. This scenario
will be studied by extracting the mass and width parameters of the FSI and comparing
them with previous measurements of the K$^*_0(700)$~\cite{Humanic:2018brf}. 
In the present Letter, a study of femtoscopic correlations with the non-identical pair combination 
$\pi^\pm$K$^0_{\rm S}$
in pp collisions at $\sqrt{s}=13$ TeV is presented for the first time
to study the nature of the K$^*_0(700)$ resonance. 
The choice of using pp collisions for this work responds to the necessity of studying the FSI in a small system in which 
the strength of the FSI is expected to be more sensitive to the system size and thus to the quark nature of the resonance~\cite{ALICE:2021ovd}. 
Due to the short-range nature of the strong interaction which might produce the resonant state, measurements in pp collisions are more suited since interparticle distances of a few fm are obtained~\cite{Fabbietti:2020bfg}. Moreover, it has already been observed that the presence of resonances in the correlation function is enhanced for measurements in small colliding systems, since the signal-to-background for the considered state scales 
as $1{\rm /multiplicity}$~\cite{Fabbietti:2020bfg}.

The results presented in this Letter 
are obtained using data collected by the ALICE Collaboration~\cite{Aamodt:2008zz,ALICE:2022wpn} during the 2015–2018 pp LHC run. The Letter is organized into seven sections: Introduction, Data Analysis, Correlation Function, Fitting, Systematic uncertainties, Results and Discussion, and Summary. The Data Analysis section gives details on how the data were taken and how the $\pi^\pm$ and 
K$^0_{\rm S}$ were reconstructed and identified. The Correlation Function section describes how the 
$\pi^\pm$K$^0_{\rm S}$ pairs were used to construct the correlation functions for this analysis. The Fitting section
describes the model used to fit the correlation functions in order to extract the source parameters and FSI parameters.
The Systematic uncertainties section discusses how the systematic uncertainties were calculated.
The Results and Discussion section presents the results for the extracted parameters and discusses their interpretation.
The Summary section summarizes the results of the present work.

\section{Data Analysis}
\label{sec:dataan}
The ALICE detector and its performance are described in detail in Refs.~\cite{Aamodt:2008zz,Alessandro:2006yt}. Collision events are selected by using the information from the V0 detectors composed of the V0C and V0A scintillator arrays~\cite{Abelev:2013vea,Abelev:2013qoq}, located on both sides of the interaction point, covering the pseudorapidity intervals $-3.7<\eta<-1.6$ and $2.8<\eta<5.1$, respectively. 
In the analysis $5\times10^8$ minimum bias triggered pp collisions at
$\sqrt{s}=13$ TeV were used.
Charged particle multiplicity classes, given in terms of multiplicity percentile intervals of the visible inelastic pp cross section, were also determined from the V0 detectors~\cite{ALICE:2020swj}.

The Time Projection Chamber (TPC)~\cite{Alme:2010ke} 
and the Inner Tracking System (ITS)~\cite{Aamodt:2008zz} were used for charged particle tracking.
These detectors cover the pseudorapidity range of $|\eta|<0.9$ and are located within a solenoid magnet with a field strength of magnitude $B = 0.5$ T. 
The momentum ($p$) determination for charged tracks was made using only the TPC space points. The ITS provided excellent spatial resolution in determining the primary collision vertex. This vertex was used to constrain the tracks reconstructed with the TPC, requiring it to be within $\pm10$ of the center of the ALICE detector.
The average momentum resolution typically obtained in this analysis for charged tracks was less 
than 10 MeV/$c$~\cite{Alessandro:2006yt}.
The selections based on the quality of track fitting~\cite{Abelev:2014ffa,Alme:2010ke,Alessandro:2006yt},
in addition to the standard track quality criteria~\cite{Alessandro:2006yt}, were used to ensure that only well-reconstructed tracks were taken into account in the analysis. 
The quality of the track was determined by the $\chi^2/N$ value for the Kalman fit to the particle trajectory in the TPC, where N is the number of TPC clusters attached to the track~\cite{Alessandro:2006yt}. The track was rejected if the value was larger than 4.0.

Analysis specific event selection criteria were also applied.
The event must have one accepted possible $\pi^{\rm \pm}$K$^0_{\rm S}$ pair.
To reduce the effects of mini-jets which tend to produce non-flat structures in the two-particle correlation functions used in femtoscopy~\cite{Acharya:2019idg}, a selection on the event transverse sphericity, 
calculated from the azimuthal distribution of tracks, was applied by requiring $S_{\rm T}>0.7$.
$S_{\rm T}$ is a scalar quantity that takes values in the range $0-1$ characterizing the event shape, 
i.e.~$S_{\rm T}\sim 0$ values represent elongated events that are ``jet-like'' and result from a single hard-scattering
of partons, whereas $S_{\rm T}\sim 1$ values
represent spherical ``non-jet-like'' events resulting from many soft parton scatterings or several hard parton scatterings. See Ref.~\cite{Acharya:2019idg} for more details.
Note that the $S_{\rm T}>0.7$ selection is estimated to have $<10\%$ effect on the multiplicity of tracks entering the
femtoscopy analysis since this selection tends to remove single hard-scattering events.
Pile-up events were rejected 
using the timing information from the V0 (for out of bunch pile-up) and multiple reconstructed vertices from tracks (or track segments in the Silicon Pixel Detector layers of the ITS)~\cite{Abelev:2014ffa,Acharya:2019idg}.
The possible effect due to remaining pile-up events passing the event selection criteria described above was investigated by performing
the analysis using only low interaction-rate data-taking
periods. No significant difference was found in the results of the analysis compared with the higher interaction-rate runs used. Both sets of runs were combined for the present analysis.

Charged particles were identified with the central barrel detectors.
Particle Identification (PID) for reconstructed tracks was carried out using both the TPC and Time-Of-Flight (TOF) detectors.
For the TPC, the specific ionization energy loss ${\rm d}E/{\rm d}x$ was measured, and for the TOF, the flight time of the particle in the pseudorapidity range $|\eta| < 0.9$ was measured~\cite{Abelev:2014ffa,Akindinov:2013tea}.
For the PID signal, a value ($N_{\sigma}$) was assigned to each track denoting the number of standard deviations between the measured PID signal and the expected values, assuming a mass hypothesis, divided by the detector resolution for both detectors
\cite{Adam:2015vja,Abelev:2014ffa,Akindinov:2013tea,Alessandro:2006yt}.
A parametrized Bethe-Bloch formula~\cite{Alessandro:2006yt} was used for the TPC PID to calculate the expected energy 
loss $\left<{\rm d}E/{\rm d}x\right>$ in the detector for a particle with a given charge, mass, and momentum. The particle mass was used to calculate the expected time-of-flight as a function of track length and momentum for the TOF PID. The detailed description of the particle identification methods is given in Ref.~\cite{Aamodt:2011zz}.

For Monte Carlo (MC) calculations, particles from pp collisions simulated by the general-purpose generator  PYTHIA8~\cite{Sjostrand:2006za} with the Monash 2013 tune~\cite{Skands:2014pea}
were transported through a GEANT3~\cite{Brun:1994aa} model of the ALICE detector.
The total number of simulated pp collisions used in this analysis is $5\times10^8$.

The methods used to select and identify individual K$^0_{\rm S}$ and $\pi^{\rm \pm}$ particles are similar to those used for the ALICE K$^{\rm \pm}$K$^0_{\rm S}$ analysis in pp collisions at 
$\sqrt{s}=13$ TeV~\cite{ALICE:2021ovd}.

K$^0_{\rm S}$ are reconstructed from their decay into $\pi^+\pi^-$, which has a branching ratio 
of $69\%$~\cite{Workman:2022ynf}. 
The neutral K$^0_{\rm S}$ decay vertices and parameters are reconstructed and calculated from pairs of detected 
$\pi^+\pi^-$ tracks, and selected based on their invariant mass and the K$^0_{\rm S}$ decay topology. The selection criteria for the K$^0_{\rm S}$
and the daughter pions are shown in Table~\ref{tab:singleKcuts}.

\begin{table}
 \centering
 \caption{$\pi^\pm$ and K$^0_{\rm S}$ selection criteria.}
 \begin{tabular}{| l | c |}
  \hline
  {\bf Neutral kaon selection} & {\bf Value} \\ \hline
  Daughter $p_{\rm T}$ & $> 0.15$ GeV/$c$ \\  
  Daughter $|\eta|$ &  $< 0.8$ \\
  Daughter DCA (3D) to primary vertex & $>0.4$ cm \\
  Daughter TPC PID [$N_\sigma$] &  $< 3$ \\
  Daughter TOF PID [$N_\sigma$] (for $p > 0.8$ GeV/$c$) &  $< 3$ \\
  Kalman fit $\chi^2/N$ & $\leq 4$ \\
  \hline
  $\left|\eta\right|$ &  $< 0.8$ \\
  DCA (3D) between daughters & $< 0.3$ cm \\
  DCA (3D) to primary vertex & $< 0.3$ cm \\
  Decay length (3D, lab frame) & $< 30$ cm \\
  Decay radius (2D, lab frame) & $> 0.2$ cm \\
  Cosine of pointing angle & $> 0.99$ \\
  Invariant mass & $0.485 < m < 0.510$ GeV/$c^{\rm 2}$ \\ \hline
  {\bf Primary pion selection} & {\bf Value} \\ \hline
  $p_{\rm T}$ & $0.15 < p_{\rm T} < 1.2$ GeV/$c$ \\
  $|\eta|$ &  $< 0.8$ \\
  Transverse DCA to primary vertex & $<2.4$ cm \\
  Longitudinal DCA to primary vertex & $<3.0$ cm \\
  TOF PID [$N_\sigma$] with valid TOF signal and $p>0.5$ GeV/$c$ & $<2$  \\
  TPC PID [$N_\sigma$] if no TOF signal for all $p$ & $< 2$  \\
  Kalman fit $\chi^2/N$ & $\leq 4$ \\
  \hline
  \end{tabular}
 
  \label{tab:singleKcuts}
\end{table}

\noindent The selection criteria are based on decay topology, i.e.~distance-of-closest-approach (DCA) between 
charged pion daughters, 
DCA of daughter pion to the primary vertex, DCA of reconstructed K$^0_{\rm S}$ to the primary vertex,
cosine of pointing angle, and decay length of K$^0_{\rm S}$, and were tuned to optimize purity and 
statistical significance.  If two reconstructed K$^0_{\rm S}$ particles share a daughter track, both are removed from the analysis. The MC samples were used to study any bias that might be induced by this 
procedure, which resulted in rejecting $<1\%$ of the K$^0_{\rm S}$ candidates~\cite{Adam:2015vja,ALICE:2021ovd}. 
Reconstructed K$^0_{\rm S}$ candidates within invariant mass range 
$0.485 < m(\pi^+\pi^-) < 0.510$ GeV/$c^2$ are used in this analysis which gives $98\pm1\%$ purity of K$^0_{\rm S}$.
The purity here is defined as signal/(signal + background). The signal and background counts are calculated by fitting a fourth-order polynomial to the side-bands of the signal region to estimate the background there
and subtracting this from the invariant mass histogram.
A Gaussian is used to fit the signal peak in the invariant mass distribution (see Fig. 2 of Ref.~\cite{Adam:2015vja}).

Primary charged pions are selected using the PID information from the TPC and TOF detectors.
The TPC is used for PID in the full momentum range, except if a valid TOF signal is available for $p>0.5$ GeV/$c$
then TOF PID is used. 
For more details, refer to Refs.~\cite{Abelev:2012sq,Adam:2015vja}.   Table~\ref{tab:singleKcuts} summarizes the criteria used for the
charged pion selection.
The average charged pion purity is found using MC simulations
to be $98.1\pm0.1\%$,
in agreement with the charged pion purity reported in Ref.~\cite{Adam:2015vja}.

Two-track effects, such as the merging of two real tracks into one reconstructed track and the splitting of one real track into two reconstructed tracks, are an important challenge for femtoscopic studies. 
A selection on the minimum separation distance between the primary pion and a daughter pion from the decay of the K$^0_{\rm S}$ in the $\pi^\pm$K$^0_{\rm S}$ pair was made from the corresponding TPC tracks using the same method as described in Ref.~\cite{ALICE:2021ovd}. 
The distance between the two tracks was calculated in different positions along their trajectory in the TPC (at radial distances from 85 to 150 cm from the interaction point) and a minimum separation distance of 20 cm was required.

\section{Measurement of correlation functions}
The momentum correlations of
$\pi^\pm$K$^0_{\rm S}$  pairs using the two-particle correlation function are studied in this analysis. The correlation
function is defined as $C(k^*)=A(k^*)/B(k^*)$, where $A(k^*)$ is the measured distribution of pairs from the same event and $B(k^*)$ is the reference distribution of pairs from mixed events. 
The denominator $B(k^*)$ is formed by mixing particles from one event with particles from 10
different events that satisfy the conditions that the
primary vertex positions along the beam direction are within 2 cm of each other, 
and have similar multiplicity, i.e. events within 2\% difference in multiplicity percentile
are mixed. Other sizes of the mixed events buffer were also investigated with no significant effect on 
the results of this work.
All events used are required to satisfy the $S_{\rm T}>0.7$ selection.
The $k^*$ is the magnitude of the momentum of each of the particles in the pair rest frame. In the present case of unequal mass particles in the pair, $m_1$ and $m_2$, $k^*$ is given by

\begin{equation}
k^*=\sqrt{\frac{a^2-m_1^2m_2^2}{2a+m_1^2+m_2^2}}
\label{kstar}
\end{equation}
where,
\begin{equation}
a\equiv (q_{\rm inv}^2+m_1^2+m_2^2)/2.
\label{kstar2}
\end{equation}

For convenience, the square of the invariant momentum difference 
$q_{\rm inv}^2=|\vec{p_1}-\vec{p_2}|^2-|E_1-E_2|^2$ is evaluated with the momenta and energies of the two particles measured in the laboratory frame. In the case where $m_1=m_2$, $k^*$ can be expressed as 
$k^*=q_{\rm inv}/2$.
A $k^*$ bin size of 
20 MeV/$c$ was used in the analyses presented in this Letter.

Correlation functions are analyzed for three cases: 1) $0-100\%$ multiplicity class and $k_T>0$ GeV/$c$, 
2) $0-100\%$ multiplicity class and $k_T < 0.5$ GeV/$c$, and 3) $0-5\%$ 
multiplicity class and $k_T < 0.5$ GeV/$c$,
where
$k_{\rm T}=|\vec{p}_{\rm T1}+\vec{p}_{\rm T2}|/2$, and where 
$\vec{p}_{\rm T1}$ and
$\vec{p}_{\rm T2}$ are the transverse momenta of the particles in the pair.
The three cases correspond to the following average $k_T$ and average charged-particle pseudorapidity 
density ($\left<dN/d\eta\right>$ in the $|\eta|<0.8$ range) values~\cite{ALICE:2020swj}:
1) $\left<k_T\right>=0.655$ GeV/$c$, $\left<dN/d\eta\right>=6.89$, 2) $\left<k_T\right>=0.323$ GeV/$c$, 
$\left<dN/d\eta\right>=6.89$, and 3) $\left<k_T\right>=0.326$ GeV/$c$, $\left<dN/d\eta\right>=21.2$.
The purpose of analyzing these cases is to obtain different femtoscopic source sizes and
to study the effect of source size on the FSI.
It has been found from femtosocpy measurements in pp collisions that the source size depends on both
$\left<k_T\right>$ and $\left<dN/d\eta\right>$~\cite{Abelev:2012ms,ALICE:2011kmy}.
In addition, case 1) was chosen to maximize the sample size and to provide a selection-free case to
compare with the other cases having multiplicity and $k_T$ selections.

\begin{figure}[]
\centering
\includegraphics[width=51mm]{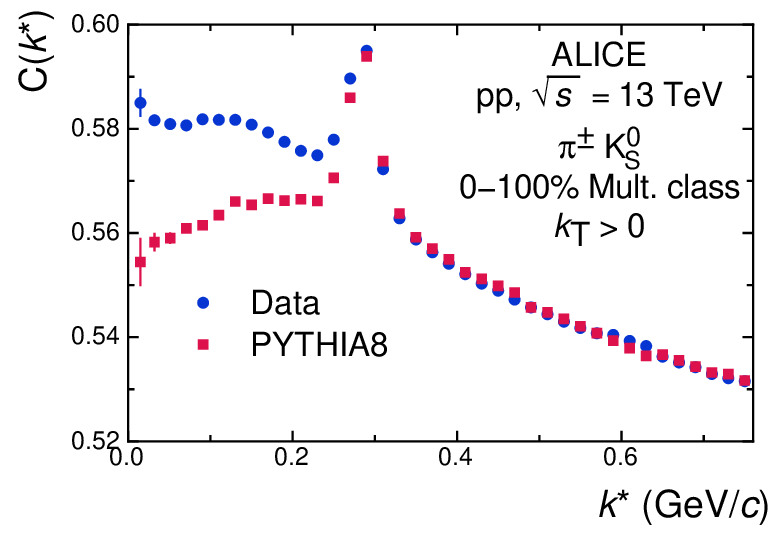}
\hspace{1mm}
\includegraphics[width=51mm]{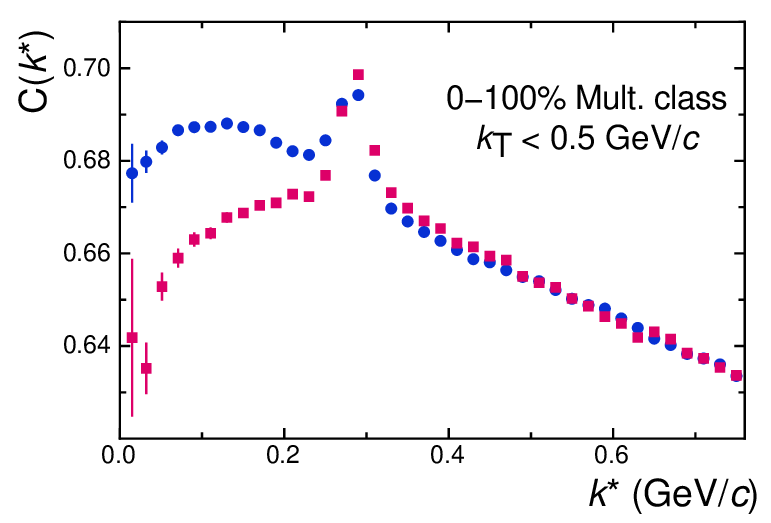}
\hspace{1mm}
\includegraphics[width=51mm]{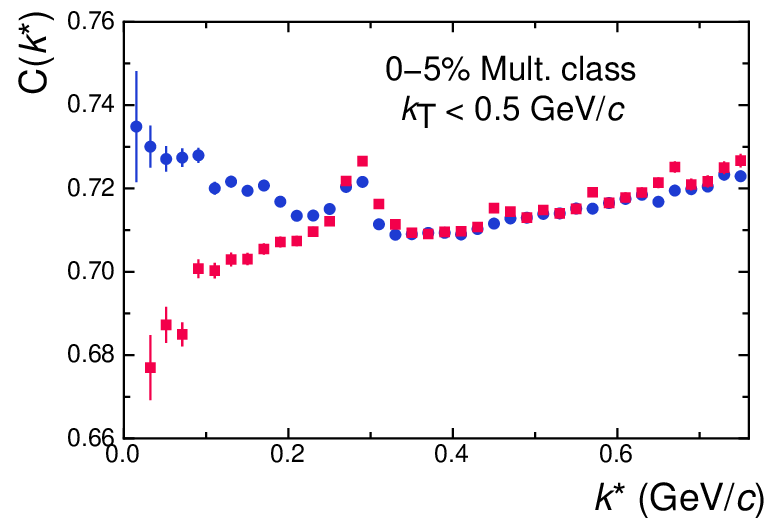}
\vspace{1mm}
\includegraphics[width=51mm]{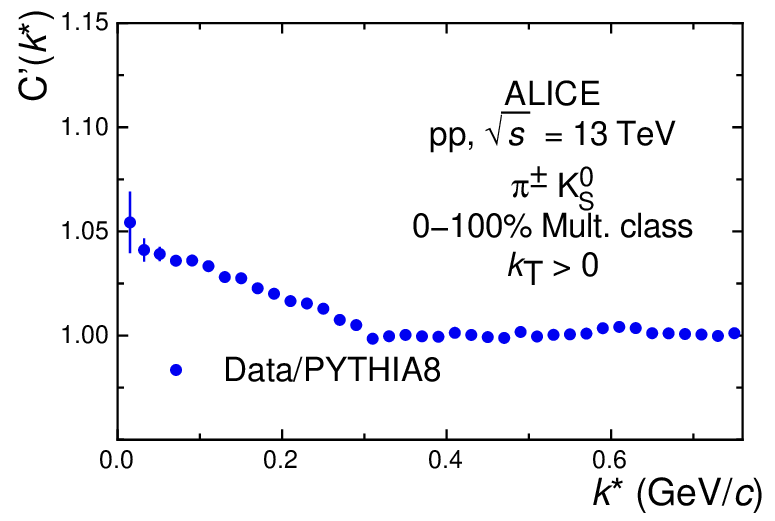}
\hspace{1mm}
\includegraphics[width=51mm]{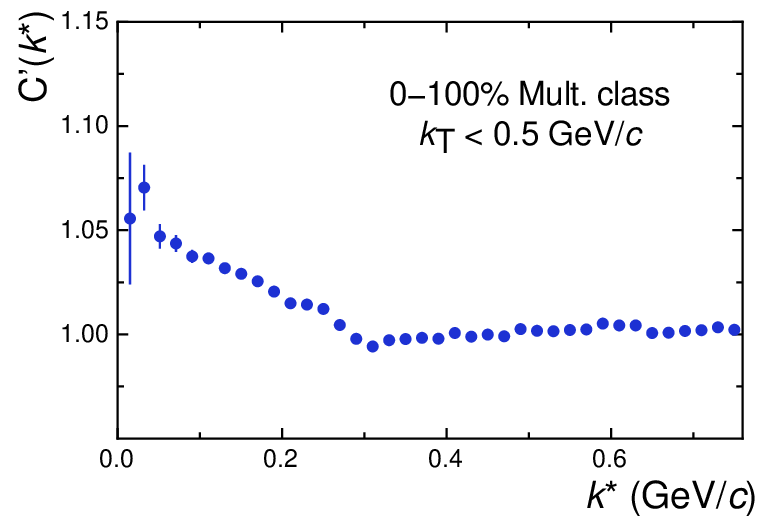}
\hspace{1mm}
\includegraphics[width=51mm]{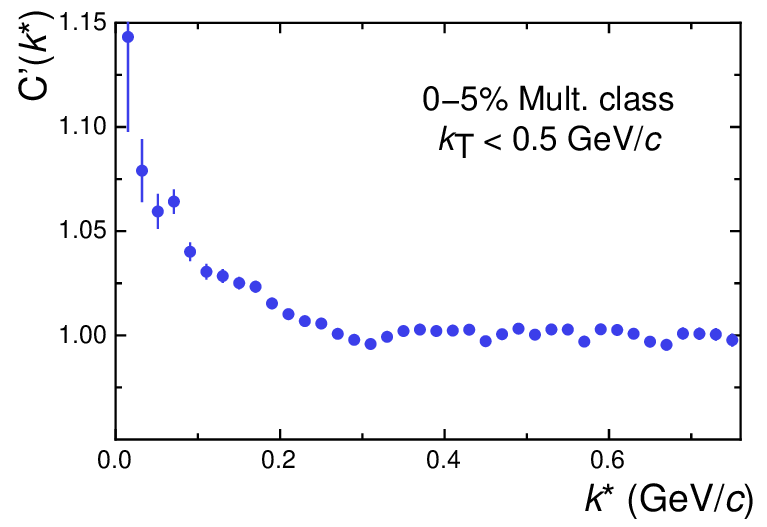}
\caption{Top row: $\pi^\pm$K$^0_{\rm S}$ correlation functions experimentally measured (blue dots) compared with PYTHIA8+GEANT3 simulations (red squares) obtained in pp collisions at $\sqrt{s}=13$ TeV 
for $0-100\%$ multiplicity class and  $k_T>0$ (left), $0-100\%$ multiplicity class and $k_{\rm T}<0.5$ GeV/$c$ (center), and 0--5$\%$ multiplicity class and $k_{\rm T}<0.5$ GeV/$c$ (right). 
The PYTHIA8+GEANT3 correlation function is normalized to
the data at $k^*=0.5$ GeV/c. Bottom row: Ratio of Data to PYTHIA8+GEANT3 simulations for the three studied cases.
Statistical uncertainties are represented by bars.}
\label{fig:raw123}
\end{figure}

Monte Carlo simulations were used to simulate correlation functions which were compared with experimental data.
Figure~\ref{fig:raw123} shows in the top row 
the correlation functions experimentally measured (blue) along with the simulated ones (red).
The MC correlation functions are normalized to the experimental ones at $k^*=0.5$ GeV/$c$ for the three cases mentioned
above. 
The single-event and mixed-event distributions of the correlation functions are summed 
over $\pi^+$K$^0_{\rm S}$ and $\pi^-$K$^0_{\rm S}$
pairs, since it is found that there is no significant difference between the 
$\pi^+$K$^0_{\rm S}$ and $\pi^-$K$^0_{\rm S}$ corresponding correlation functions.
The decay of the K$^*(892)$ meson is clearly seen at $k^*\sim 0.3$ GeV/$c$ for all cases.
For $k^*>0.35$ GeV/$c$ a non-flat baseline is also observed in all cases. 
This non-flat baseline is associated with soft parton fragmentation, or mini-jets, that are not completely suppressed by the transverse sphericity selection~\cite{ALICE:2011kmy,ALICE:2019gcn,ALICE:2021cpv}, as well as the presence of momentum conservation effects. Non-flat baselines in two-particle correlation functions obtained in pp collisions are often observed~\cite{Acharya:2018kpo,ALICE:2011kmy,ALICE:2019gcn}. In particular, measured correlation functions
show a decreasing dependence of
the baseline with increasing $k^*$ for low multiplicity classes, and a reversal of this dependence for higher multiplicity
classes, as seen in Fig.~\ref{fig:raw123}.
This effect in the data is seen to be present as well in the PYTHIA8 simulations.
The simulations well reproduce the K$^*(892)$ peak and the background visible at larger $k^*$, hence in order to remove
these two contributions, the measured correlation function is subsequently divided by the simulated one,
defined as $C~'(k^*)$, as shown in
the bottom panels of Fig.~\ref{fig:raw123}. The statistical
uncertainty from the MC correlation function is propagated with the uncertainty from the experimental one in the ratio, which becomes the final correlation function.

Finite track momentum resolution can smear the relative momentum correlation functions used in this analysis. This
effect is corrected using MC simulations as done in previous works~\cite{Aamodt:2008zz,ALICE:2022wpn}.
It is found that
the effect of the momentum resolution correction is small for the very lowest $k^*$ bin with the largest statistical error bars and negligible for the rest of the bins, resulting in a $<3\%$ effect on the extracted fit parameters.

\section{Fitting}
The momentum resolution corrected ratio of the experimental $\pi^\pm$K$^0_{\rm S}$ correlation function to the MC correlation function was fitted
by a model in order to extract information on the size of the source, as well as the strength and nature
of the FSI between the particles in the pair. The fit function is given by,

\begin{equation}
C~'(k^*) = \kappa\left[C_{\rm Lednicky}(k^*)+\epsilon \frac{dN_{BW}}{dm}\frac{dm}{dk^*}\right]
\label{eq:fit5}
\end{equation}

where,
\begin{equation}
\frac{dN_{BW}}{dm}\propto \frac{\Gamma_{892}}{(m-m_{892})^2+\Gamma_{892}^2/4}
\label{eq:fit6}
\end{equation}
is the Breit--Wigner resonance distribution. This last term fits out any residual presence of the K$^*(892)$ peak
(see below).

The quantities $\epsilon$ and $\kappa$, where $\epsilon$ is the magnitude of a 
correction term on the MC modeling of the K$^*(892)$ (see below) and
$\kappa$ is an overall normalization factor, are fit parameters, and $\Gamma_{892}$ and $m_{892}$ are the full-width at half maximum (FWHM) and mass
of the K$^*(892)$, respectively, taken from the Review of Particle Physics~\cite{Workman:2022ynf}. 
The first term in Eq.~\ref{eq:fit5} is a modified version of the Lednicky
parametrization~\cite{Abelev:2006gu,Lednicky:1981su,Lednicky:2005af} which assumes that the 
pair interaction is due to strong 
final-state interaction of a near-threshold resonance.
The second term in Eq.~\ref{eq:fit5} is used to fit out the small residual bump in the ratio that results from a slight overcompensation of the MC in modeling the K$^*(892)$ peak in the data that can be seen in
Fig.~\ref{fig:raw123}, located at $k^*\sim0.3$ GeV/$c$. Fitting out this residual bump results in an improved
$\chi^2/{\rm ndf}$ for all of the fits.

A Gaussian distribution of the source size in the
pair reference frame is assumed in the FSI parameterization. 
More general forms for this distribution could be used, but using the Gaussian results in the analytic form 
of the Lednicky equation. Another motivation for staying with the Gaussian is to facilitate comparisons with
previous published results that also used the Gaussian distribution.

The quantity $C_{\rm Lednicky}(k^*)$ has the form

\begin{equation}
C_{\rm Lednicky}(k^*)=1+\left( \frac{\lambda\alpha}{2}\right)\left[\left|\frac{f(k^*)}{R}\right|^2+\frac{4\mathcal{R}f(k^*)}{\sqrt{\pi}R}F_1(2k^* R)-\frac{2\mathcal{I}f(k^*)}{R}F_2(2k^* R)+\Delta C\right]
\label{eq:fit1}
\end{equation}
and
\begin{equation}
F_1(z)=\int^{z}_{0}dx\frac{e^{x^2-z^2}}{z}; ~~~~F_2(z)=\frac{1-e^{-z^2}}{z}.
\label{eq:fit2}
\end{equation}

 $\alpha$ is the symmetry parameter and is set to 0.5 assuming symmetry in K$^0$ and 
$\overline{\rm K^0}$ production since the K$^0_{\rm S}$ is a linear combination of these; $R$ is the radius parameter
of the source; and $\lambda$ is the correlation strength.
The term $f(k^*)$ is the s-wave $\pi^\pm$K$^0_{\rm S}$ scattering amplitude whose FSI contribution is the near-threshold resonance.
A relativistic Breit--Wigner amplitude is assumed,
\begin{equation}
f(k^*) = \frac{\gamma}{M_R^2-s-i\gamma k^*}.
\label{eq:fit3}
\end{equation}

 In Eq.~\ref{eq:fit3}, $M_R$ is the mass of the resonance, and 
$\gamma$ is the coupling of the resonance to its decay channel, i.e. $\pi^\pm$K$^0_{\rm S}$ . Also, $s=(\sqrt{m_{\rm K}^2+k^{*2}}+\sqrt{m_{\pi}^2+k^{*2}})^2$ is the square of the energy of the pair in its rest frame.
A Breit--Wigner form was chosen for $f(k^*)$ since the fitted $M_R$ and
$\gamma$ to the FSI resonance from the present work will be compared with other measurements that used 
the Breit--Wigner form in order to identify the resonance~\cite{BES:2010soq,E791:2002xlc}.

The quantity $\Delta C$ is a
correction to the derivation of Eq.~\ref{eq:fit1}, that assumes spherical outgoing waves, to account for the true scattered waves in the inner region of the short-range potential~\cite{Abelev:2006gu,ALICE:2021ovd}, and
is given by,

\begin{equation}
\Delta C = \frac{(2+m_\pi/m_{\rm K}+m_{\rm K}/m_\pi)}{2\sqrt{\pi}R^3\gamma} |f(k^*)|^2.
\label{eq:fit4}
\end{equation}

As a test, a p-wave term was added to the s-wave term in the scattering amplitude in deriving the Lednicky equation to study whether there was interference of the K$^*(892)$ with the s-wave FSI. It was found that the p-wave term had a negligible effect on the fits, and was thus ignored. 

The fitting strategy was to make a six-parameter fit of Eq.~\ref{eq:fit5} to the corrected
ratio of the experimental $\pi^\pm$K$^0_{\rm S}$ correlation
function to the corresponding MC correlation function to extract $R$, $\lambda$, $M_R$, $\gamma$, 
$\epsilon$, and $\kappa$. The nominal fit range
is $0<k^*<0.76$ GeV/$c$ in all cases.
The nominal maximum of 0.76 GeV/$c$ of the fit range was set to give the optimal overlap between the experimental and MC correlation functions in the baseline region.

\begin{figure}[]
\centering
\includegraphics[width=51mm]{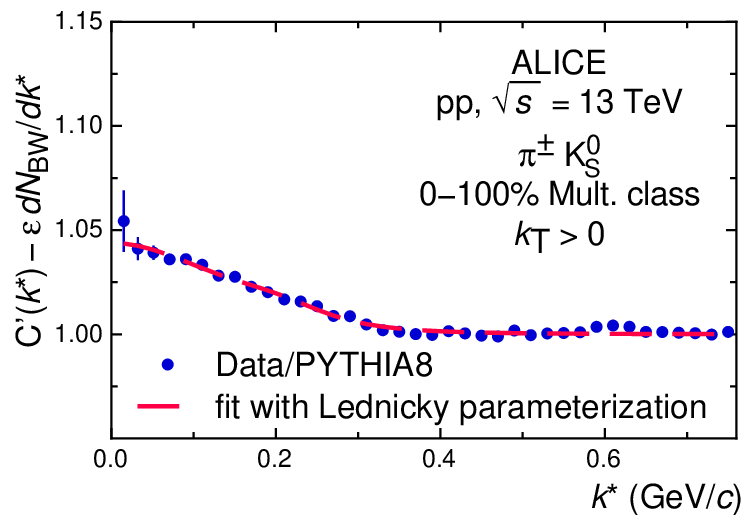}
\hspace{1mm}
\includegraphics[width=51mm]{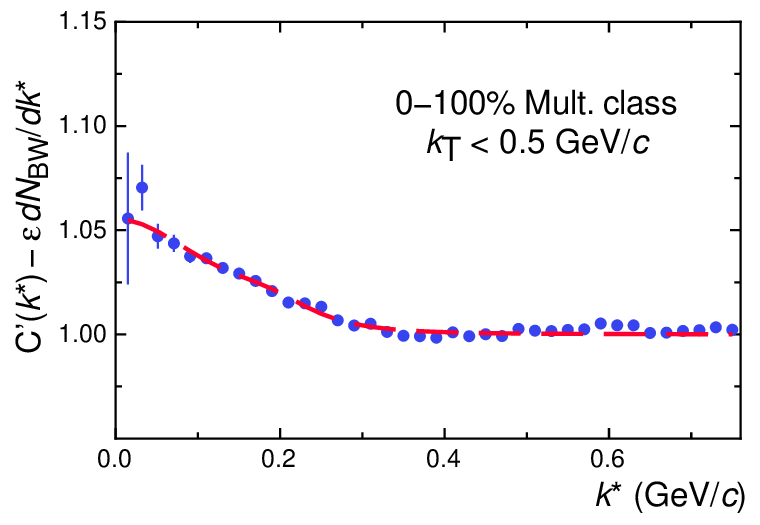}
\hspace{1mm}
\includegraphics[width=51mm]{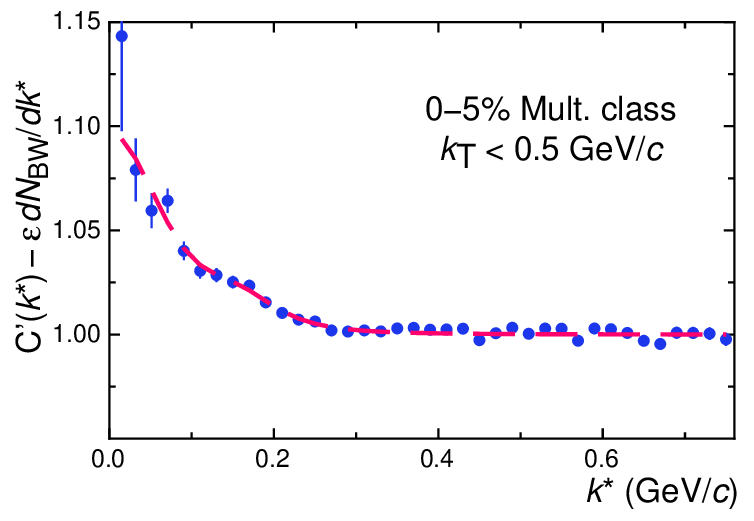}
\caption{Example fit of Eq.~\ref{eq:fit1} to the corrected correlation functions after Eq.~\ref{eq:fit5} has been used to remove the PYTHIA8+GEANT3 overcompensation of the
K$^*(892)$, for $\pi^\pm$K$^0_{\rm S}$ from $\sqrt{s}=13$ TeV pp collisions
for $0-100\%$ multiplicity class and  $k_T>0$ (left), $0-100\%$ multiplicity class and $k_{\rm T}<0.5$ GeV/$c$ (center), and 0--5$\%$ multiplicity class and $k_{\rm T}<0.5$ GeV/$c$ (right). Statistical uncertainties are represented as bars.}
\label{fig:fit123a}
\end{figure}

Figure~\ref{fig:fit123a} shows the correlation
functions and fits. The MC overcompensation of the K$^*(892)$ has been removed from the ``Data/MC'' points by subtracting out the second term in Eq.~\ref{eq:fit5} in order to show how well $C_{\rm Lednicky}(k^*)$ fits the ratio,
and the ratio has been divided by $\kappa$.
The $\chi^2/$ndf for the fits shown in Fig.~\ref{fig:fit123a} are 1.6, 1.8, and 0.92, with p-values of
1.7\% and 0.36\% and 60\%, respectively.

\section{Systematic uncertainties}
Table~\ref{tab:fitresults} shows the total systematic uncertainties on the $R$, $\lambda$
$M_R$, and $\gamma$
parameters extracted from the $\pi^\pm$K$^0_{\rm S}$ correlation function in pp collisions at $\sqrt{s}=13$ TeV.

The ``fit systematic uncertainty'' column reports the systematic uncertainty due
to varying the $k^*$ fit range. Varying the fit range by $20\%$ resulted in $<3\%$
effect on the fit parameters. 
Fitting uncertainties were calculated including correlations among the fit parameters as done using a MINOS algorithm
in order to obtain conservative estimates of the uncertainties~\cite{james1971}.

The ``selection systematic uncertainty'' column reports the systematic uncertainty related to the
variation of track and PID selection criteria used in the data analysis.
To determine this, the single particle selection criteria shown in Table~\ref{tab:singleKcuts} were varied by $\pm10$\%, and the value chosen for the minimum separation distance of same-sign tracks was varied by $\pm20$\%~\cite{ALICE:2021ovd}. 
The systematic uncertainty related to the sphericity selection of $S_{\rm T}>0.7$ is also included in this source
of systematic uncertainty, where $S_{\rm T}$ was
varied by $\pm 10\%$ from its nominal selection value.
The uncertainty was estimated from the variation of the results with respect to those obtained with the nominal selections. The resulting relative systematic uncertainties are of about
$10\%$ for $\lambda$, about $5\%$ for $R$, and about $2\%$ for the other parameters. 

The ``total systematic uncertainty'' column is obtained as the sum in quadrature of the contribution of the two sources described above. The ``total uncertainty'' column is the sum in quadrature of the statistical uncertainty
and the total systematic uncertainty.
As seen, the total systematic uncertainties tend to be greater than or comparable to the statistical uncertainties.
Table~\ref{tab:syspercent} shows an approximate breakdown of the relative systematic uncertainties (in percentage)
from the different variations considered. See Table~\ref{tab:singleKcuts} in Section~\ref{sec:dataan} for the nominal values of the selection criteria. Note that ``min. sep. var.'' refers to the variation of the selection for minimum separation between
K$^0_{\rm S}$ daughter and primary pions in the TPC, mentioned earlier, and ``$m(\pi^+\pi^-)$ and primary vertex variations'' refer to
the combined effect of varying the invariant mass selection for K$^0_{\rm S}$ and varying the selection for the primary vertex
of the event.
As seen, in general the variations have the largest effect on $\lambda$ and the smallest effect on $M_R$ and $\gamma$, with the $S_{\rm T}$ variation having the largest single-variation effect on all of the parameters.

\section{Results and discussion}
The $R$, $\lambda$,
$M_R$, and $\gamma$
parameters extracted from the present analysis of $\pi^\pm$K$^0_{\rm S}$ correlation functions in
pp collisions at  $\sqrt{s}=13$ TeV
are reported in Table~\ref{tab:fitresults} for the three cases mentioned above. The $\lambda$ parameters
are corrected for purity by dividing the extracted $\lambda$ values with the product of the $\pi^\pm$
and K$^0_{\rm S}$ purities (see Section~\ref{sec:dataan}).

\begin{table}
 \centering
  \caption{Fit results for R, $\lambda$, $M_R$, and $\gamma$ showing statistical and systematic uncertainties
  from the present analysis. Uncertainties are symmetric unless specified otherwise.
  See the text for the description of the various sources of uncertainties.}
 \begin{tabular}{| c | c | c | c | c | c | c | c |}
  \hline
{$R$, $\lambda$, $M_R$, or $\gamma$} & fit & statistical & fit & selection & total & total \\ 
  & value  & uncertainty  & systematic  & systematic  & systematic  & uncertainty  \\ 
  &      &     &  uncertainty   & uncertainty  &  uncertainty   &  \\ \hline \hline
$0-100\%$  \\ 
multiplicity class \\ 
$k_T>0$ \\ \hline
{$R$ (fm)} & 0.912 & 0.037 & 0.011 & 0.053 & 0.054 & 0.065 \\ \hline
{$\lambda$} & 0.0783 & +0.0096 & 0.0032 & 0.0078 & 0.0084 & +0.0127 \\ 
    &   & -0.0086   &   &    &    &-0.0121  \\ \hline 
{$M_R$ (GeV/$c^2$)} & 0.833 & 0.002 & 0.006 & 0.013 & 0.015 & 0.015 \\ \hline
{$\gamma$ (GeV)} & 0.890 & 0.015 & 0.012 & 0.016 & 0.020 & 0.025 \\ \hline \hline
$0-100\%$  \\
multiplicity class \\
$k_T<0.5$ GeV/$c$ \\ \hline 
{$R$ (fm)} & 1.063 & 0.058 & 0.015 & 0.064 & 0.066 & 0.088 \\ \hline
{$\lambda$} & 0.111 & 0.017 & 0.004 & 0.013 & 0.014 & 0.022 \\ \hline
{$M_R$ (GeV/$c^2$)} & 0.804 & 0.003 & 0.005 & 0.013 & 0.014 & 0.014 \\ \hline
{$\gamma$ (GeV)} & 0.801 & 0.023 & 0.020 & 0.014 & 0.024 & 0.033 \\ \hline \hline
$0-5\%$  \\
multiplicity class \\
$k_T<0.5$ GeV/$c$ \\ \hline
{$R$ (fm)} & 1.618 & +0.136 & 0.015 & 0.089 & 0.090 & +0.163 \\ 
    &   & -0.109   &   &    &    &-0.142  \\ \hline 
{$\lambda$} & 0.274 &+0.077 & 0.001 & 0.026 & 0.026 & +0.081 \\ 
    &   & -0.053   &   &    &    & -0.059 \\ \hline 
{$M_R$ (GeV/$c^2$)} & 0.765 & 0.004 & 0.002 & 0.012 & 0.013 & 0.013 \\ \hline
{$\gamma$ (GeV)} & 0.714 & +0.042 & 0.005 & 0.013 & 0.014 & +0.044 \\
   &    & -0.037            &   &   &   & -0.039    \\ \hline
  \end{tabular}
 
  \label{tab:fitresults}
\end{table}

\begin{table}
 \centering
   \caption{Breakdown of the relative systematic uncertainties for R, $\lambda$, $M_R$, and $\gamma$ from the variation of track, PID and mixed-event selection criteria. The $\%\Delta$ row is the percentage that the quantity was changed.
  See the text for the description of the various uncertainties.}
 \begin{tabular}{| c | c | c | c | c | c | c | c | c |}
  \hline
 {Quantity}& Fit  & Min. & TOF, TPC & DCA & $m(\pi^+\pi^-)$ and  & Multiplicity & Decay & $S_{\rm T}$ \\ 
  {changed}& range   & sep.  & $N_\sigma$  & var.  & primary vertex  & difference for & length & var. \\ 
  { }          &       & var.     &         &         & var. & event mixing  &     &  \\
  \hline \hline
  {$\% \Delta$} &  $20$    &   $20$  & $10$  &  $10$   & $10$ & $10$ & $10$ & $10$\\  \hline \hline
{$\%R$} & 1  & 1 & 2 & 2 & 1 & 1 & 1 & 3\\ \hline
{$\%\lambda$} & 3  & 5 & 3 & 3 & 3 &2 & $2$ & 5 \\  \hline
{$\%M_R$} & 1 &  $<1$ & $<1$ & $<1$ & $<1$ & $<1$ & $<1$ & 2 \\ \hline
{$\%\gamma$} & 2  & 1 & $<1$ & $<1$ & $<1$ & $<1$ & $<1$ & 2 \\ \hline
  \end{tabular}

  \label{tab:syspercent}
\end{table}

Since the main goal of this measurement is to study the K$^*_0(700)$ resonance, one must first establish 
that the FSI of the $\pi^\pm$K$^0_{\rm S}$ pair occurs indeed through this resonance.
This can be done by comparing the measured $M_R$ and $\gamma$ parameters extracted from this analysis with
previously measured values of $M_R$ and $\Gamma_R$ for the K$^*_0(700)$~\cite{BES:2010soq,E791:2002xlc}, where $\Gamma_R$
is the FWHM of the relativistic Breit--Wigner resonance distribution, whose amplitude is expressed as~\cite{Bohm:2004zi},

\begin{equation}
f(s) \sim \frac{1}{M_R^2-s-iM_R \Gamma_R}.
\label{eq:rbw}
\end{equation}

Comparing this denominator with the denominator of Eq.~\ref{eq:fit3}, one can obtain an estimate for $\Gamma_R$ from the
present results,

\begin{equation}
\Gamma_R = \frac{\left<k^*\right>\gamma}{M_R},
\label{eq:gamma}
\end{equation}

where $\left<k^*\right>$ is the average of $k^*$ determined by weighting $k^*$ by the experimental
$dN/dk^*$ distribution over the fit range used in fitting Eq.~\ref{eq:fit5} to the
correlation function. Table~\ref{tab:gamma} lists the values of $\Gamma_R$ extracted from the
present work using Eq.~\ref{eq:gamma} for the three cases studied. The uncertainties shown for
$\left<k^*\right>$ are estimated by
considering different $k^*$ ranges for calculating the average, namely $0<k^*<0.6$ GeV/$c$ and $0<k^*<2$ GeV/$c$, 
and taking the differences from the 
nominal $\left<k^*\right>$ 
to obtain conservative estimates of the uncertainties.

\begin{table}
\centering
 \caption{The $\left<k^*\right>$ and corresponding $\Gamma_R$ extracted from the three cases
 measured in the present work using Eq.~\ref{eq:gamma}.}
 \begin{tabular}{| c | c | c |}
  \hline
Case & $\left<k^*\right>$ (GeV/$c$)  & $\Gamma_R$ (GeV/$c^2$) \\ \hline
 $0-100\%$ multiplicity class, $k_T>0$  &$0.403^{+0.093}_{-0.056}$ & $0.430^{+0.088}_{-0.053}$ \\  \hline
 $0-100\%$ multiplicity class, $k_T<0.5$ GeV/$c$ & $0.408^{+0.060}_{-0.050}$& $0.406^{+0.050}_{-0.042}$ \\ \hline
 0 -- 5\% multiplicity class, $k_T<0.5$ GeV/$c$ & $0.418^{+0.072}_{-0.053}$ & $0.390^{+0.068}_{-0.051}$ \\ \hline
 \end{tabular}

 \label{tab:gamma}
\end{table}

\begin{figure}
	\centering
		\includegraphics[scale=.7]{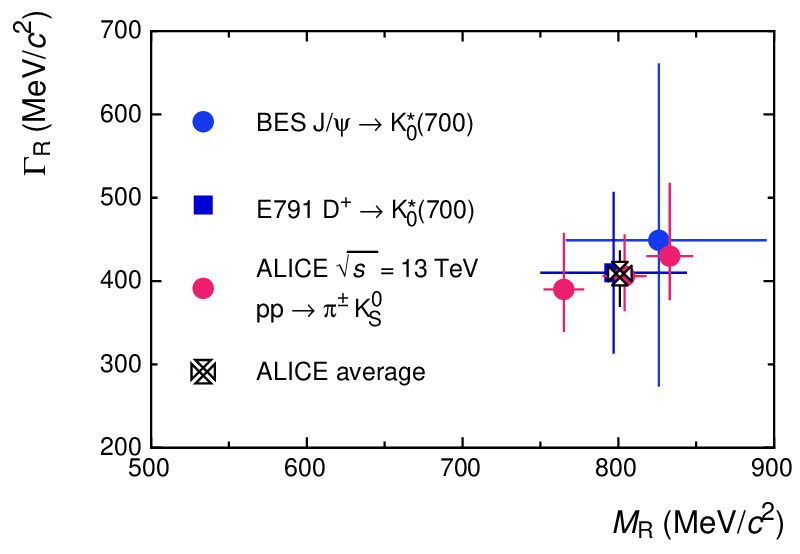}
	\caption{The extracted Breit--Wigner parameters from the $\pi^\pm$K$^0_{\rm S}$ femtoscopic correlation in
pp collisions at  $\sqrt{s}=13$ TeV
 compared with
	those for K$^*_0(700)$ from the BES~\cite{BES:2010soq} and the E791~\cite{E791:2002xlc} experiments. 
	The horizontal and vertical bars represent the total uncertainties.
	The ``ALICE average'' value is the weighted average
	of the three ALICE points.}
	\label{fig:M_Gamma}
\end{figure} 

Figure~\ref{fig:M_Gamma} compares the values of $M_R$ and $\Gamma_R$ extracted in the
present work with measurements of these quantities for the K$^*_0(700)$ from the 
BES~\cite{BES:2010soq}
and E791 Collaborations~\cite{E791:2002xlc}. The BES Collaboration measured the relativistic Breit--Wigner parameters of the K$^*_0(700)$ through the decay of the
J/$\psi$ meson, whereas the E791 Collaboration measured them through the decay of the D$^+$ meson.
The total uncertainties defined as the quadratic sum of the statistical and systematic uncertainties are shown on the points for all cases.
As seen,
the values reported in this work agree within uncertainties with the 
K$^*_0(700)$ Breit--Wigner parameters measured in the other two experiments.
It is seen that the present results have smaller uncertainties than the previous measurements.
It is also seen that although the three $\Gamma_R$ values from the present work agree within uncertainties, the differences among the three $M_R$ values are outside of their uncertainties.
This could be a consequence of using the Breit--Wigner function to fit a resonance where the condition
$\Gamma_R\ll M_R$ is not fulfilled,
which can lead to kinematic dependences on the extracted $M_R$ and 
$\Gamma_R$~\cite{Workman:2022ynf,Bohm:2004zi}. However, these differences in $M_R$ are small compared with
the extracted $M_R$ values, and thus it is judged that
these results strongly support the assumption that the resonance responsible for the FSI of the 
$\pi^\pm$K$^0_{\rm S}$ pairs
studied in the present work is the K$^*_0(700)$ resonance.

\begin{figure}
	\centering
		\includegraphics[scale=.7]{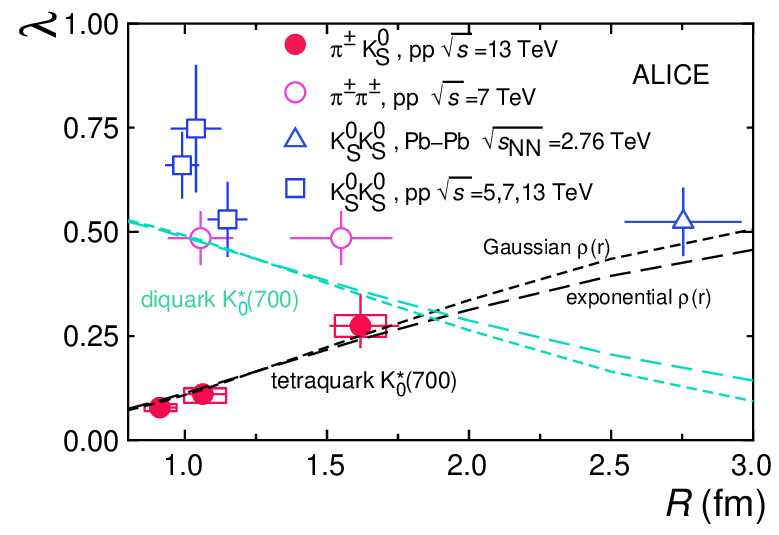}
	\caption{
The $\lambda$ parameter as a function of source size $R$ extracted from the $\pi^\pm$K$^0_{\rm S}$ femtoscopy measurement in pp collisions at $\sqrt{s}=13$ TeV. Results are compared with the previous ALICE measurements, obtained from K$^0_{\rm S}$K$^0_{\rm S}$~\cite{Adam:2015vja,ALICE:2021ovd} and $\pi\pi$~\cite{ALICE:2011kmy}  femtoscopy studies in pp and Pb–Pb collisions and the calculations from a toy geometric model (see text). The model
calculations for the tetraquark and diquark hypotheses for the K$^*_0(700)$ are shown as black and light green dashed lines, respectively,
the short dashed lines representing the Gaussian $\rho(r)$ and the long dashed lines representing the exponential 
$\rho(r)$. }
	\label{fig:R_lam_toy}
\end{figure}

The extracted $R$ and $\lambda$ parameters shown in Table~\ref{tab:fitresults} can be used
to obtain information about the quark configuration of the K$^*_0(700)$. Figure~\ref{fig:R_lam_toy}
compares the values of $R$ and $\lambda$ extracted in the present work with 
published results
for these parameters from ALICE measurements in pp and Pb--Pb collisions in which $\pi\pi$ and 
K$^0_{\rm S}$K$^0_{\rm S}$ pairs were analyzed~\cite{Abelev:2012ms,Adam:2015vja,ALICE:2021ovd,ALICE:2011kmy}. 
The $\pi^\pm$K$^0_{\rm S}$ results are shown with separate statistical (error bars) and systematic (boxes) uncertainties, whereas for the previous results, the error bars represent the combination of the statistical
and systematic uncertainties.
For the
$\pi\pi$ femtoscopic measurements in pp collisions at $\sqrt{s}=7$ TeV reported in~\cite{ALICE:2011kmy}
with average $k_{\rm T}$ values of $\sim 0.15$ and $\sim 0.35$ GeV/$c$, the $\lambda$
values are given as varying in the range $0.42-0.55$, so $\lambda$ is plotted as the center
of this range with uncertainties extending to the upper and lower limits of the range. 

For the $R$ parameter, the values from the present $\pi^\pm$K$^0_{\rm S}$ analysis
are comparable with the published $\pi\pi$ and K$^0_{\rm S}$K$^0_{\rm S}$
measurements in pp collisions, i.e.~in the range 1--2 fm, as would be expected from pp collisions where the source size is $\sim1$ fm.
For the $\lambda$ parameter, whereas the results from $\pi\pi$ and K$^0_{\rm S}$K$^0_{\rm S}$
are compatible with values of about $0.5$ or greater, for the present $\pi^\pm$K$^0_{\rm S}$ analysis
significantly lower values are obtained, ranging from about 0.05 to about 0.25 depending on $R$.
The expectation is that
$\lambda$ would be the same for $\pi^\pm$K$^0_{\rm S}$ as for the identical-meson measurements.
The $\lambda$ value of $\sim 0.5$ has been shown to be due to the presence of long-lived resonances
whose decay into the detected mesons impacts the measurement of the ``direct''  mesons coming from the source of interest~\cite{ALICE:2021ovd,Humanic:2008nt}.
Another significant difference between the
present $\pi^\pm$K$^0_{\rm S}$ results and the $\pi\pi$ and K$^0_{\rm S}$K$^0_{\rm S}$ results
is that $\lambda$ has a strong $R$ dependence for the former, whereas there is no
significant dependence of $\lambda$ on $R$ for the latter, i.e. even extending $R$ to the value from Pb--Pb collisions shows no significant effect on $\lambda$.

As discussed in
Refs.~\cite{ALICE:2021ovd} and~\cite{Acharya:2018kpo}, a physics effect that could
cause this difference in $\lambda$ values for $\pi^\pm$K$^0_{\rm S}$ pairs is related to the 
possibility that the K$^*_0(700)$ resonance, 
that is assumed to be solely responsible for
the FSI in the $\pi^\pm$K$^0_{\rm S}$ pair, is actually a tetraquark state of the form
$({\rm q_1},\overline{\rm q_2}, {\rm q_3}, \overline{\rm q_3})$, in which ${\rm q_1}$, ${\rm q_2}$ and ${\rm q_3}$ indicate the flavor of the valence quarks
of the $\pi$ and K$^0_{\rm S}$. In particular, ${\rm q_1}$ and ${\rm q_2}$ can be a u or s quark, while ${\rm q_3}$ is a
d quark.
For example, the quark
content of a tetraquark K$^*_0(700)^+$ would be u$\overline{\rm s}$d$\overline{\rm d}$, whereas
the diquark version would be u$\overline{\rm s}$.
The strength of
the FSI through a tetraquark K$^*_0(700)^+$ could be decreased by the small source size of 
the $\pi^\pm$K$^0_{\rm S}$
source, i.e.~at $R\sim 1$ fm as is measured in these collisions. 
This could occur since d$-\overline{\rm d}$ annihilation
would be enhanced due to the proximity of the $\pi^\pm$ and K$^0_{\rm S}$ at their creation,
which would open up a non-resonant channel in the scattering process that would be reflected by reducing $\lambda$.
For a FSI through a diquark K$^*_0(700)^+$, 
with the form u$\overline{\rm s}$, the small source geometry should not reduce its strength.
For the K$^0_{\rm S}$K$^0_{\rm S}$ and $\pi\pi$ cases, $\lambda$ should not be affected by
the source size since the pair correlation is
dominated by the effect of quantum statistics, for which in the ideal case $\lambda$ does not depend on $R$, 
and which is found to be much stronger than the strong FSI present
for these identical particle pairs~\cite{Abelev:2006gu}. 

In order to demonstrate the $R$ dependence of $\lambda$ for a tetraquark or a diquark K$^*_0(700)$ based on
the geometric considerations discussed above, a simple toy model is constructed, taking the form of the
$\lambda$ factor for a tetraquark state,

\begin{equation}
\lambda=\lambda_0(1-aP)
\label{eq:model1}
\end{equation}

and for a diquark,

\begin{equation}
\lambda=\lambda_0aP
\label{eq:model2}
\end{equation}

where,

\begin{equation}
P\equiv\frac{\int \rho(r)\rho(|\vec{r}-\vec{R}|) dV }{\int |\rho(r)|^2 dV}
\label{eq:model3}
\end{equation}
can be considered the ``overlap probability'' between the $\pi$ and K$^0_{\rm S}$ in the pair as they are emitted from the
pp collision. The quantity $\rho(r)$ is the meson volume distribution, assumed to be the same for the $\pi$ and K$^0_{\rm S}$, $\lambda_0$ is the maximum value for $\lambda$, and $a$ is essentially the ``d$-\overline{\rm d}$ annihilation efficiency'' that in principle could take any value in the range $0 - 1$.
Assuming $\rho(r)\sim e^{-r^2/(2\sigma^2)}$ or $\sim e^{-r/r_0}$, $\lambda_0=0.6$, the average value for $\pi\pi$
and K$^0_{\rm S}$K$^0_{\rm S}$ measurements from Refs.~\cite{ALICE:2011kmy} and ~\cite{Adam:2015vja,ALICE:2021ovd}, and assuming $100\%$  d$-\overline{\rm d}$ annihilation efficiency
for any non-zero overlap, $a=1$, 
the free parameters of the model, i.e. $\sigma$ and $r_0$, are adjusted to give a good fit to the
$\pi^\pm$K$^0_{\rm S}$ measurements.
The results from Eqs.~\ref{eq:model1} and~\ref{eq:model2} are shown 
in Fig.~\ref{fig:R_lam_toy}, along with the results from $\pi^\pm$K$^0_{\rm S}$ measurements of this work
and published ALICE measurements for K$^0_{\rm S}$K$^0_{\rm S}$~\cite{Adam:2015vja,ALICE:2021ovd} and 
$\pi\pi$ pairs~\cite{ALICE:2011kmy} from pp and Pb--Pb collisions. 
The free model parameters are set to $\sigma=1.1$ fm and $r_0=0.85$ fm for the Gaussian (short dashed lines) and exponential (long dashed lines)
distributions, respectively, which are considered reasonable values since hadronic sizes are expected to be $\sim 1$ fm.
As seen, using reasonable
model parameter values, the tetraquark case, Eq.~\ref{eq:model1}, describes the $R$ dependence of $\lambda$ from the present measurements
well for both the Gaussian and exponential meson shapes as being a geometric effect. The diquark case is seen
to predict an $R$ dependence that is incompatible with the measured one.

Therefore, the present results of $\pi^\pm$K$^0_{\rm S}$ 
femtoscopy in pp collisions at $\sqrt{s}=13$ TeV 
suggest that the K$^*_0(700)$ is a tetraquark state.

\section{Summary}
Femtoscopic correlations with the particle pair combination $\pi^\pm$K$^0_{\rm S}$
are studied in pp collisions at $\sqrt{s}=13$ TeV for the first time by the 
ALICE experiment at the LHC. Source parameters and final-state interaction parameters 
are extracted by fitting a model
based on a Gaussian distribution of the source to the experimental two-particle correlation functions. The model used 
assumes that solely the final-state interaction through a resonance determines the correlations, and is defined
in terms of a mass and the coupling parameter to the decay into
a  $\pi^\pm$K$^0_{\rm S}$ pair. 
The extracted mass and width parameters of the FSI are consistent with previous measurements of 
the K$^*_0(700)$ resonance,
and the smaller value and increasing behavior of the $\lambda$ parameter with $R$
compared with identical boson measurements give support that the
K$^*_0(700)$ is a four-quark state, i.e a tetraquark
state~\cite{Guo:2013nja}.
A simple geometric model that assumes a tetraquark FSI describes well the
$R$ dependence of $\lambda$ extracted from the measured correlation functions.


\newenvironment{acknowledgement}{\relax}{\relax}
\begin{acknowledgement}
\section*{Acknowledgements}

The ALICE Collaboration would like to thank all its engineers and technicians for their invaluable contributions to the construction of the experiment and the CERN accelerator teams for the outstanding performance of the LHC complex.
The ALICE Collaboration gratefully acknowledges the resources and support provided by all Grid centres and the Worldwide LHC Computing Grid (WLCG) collaboration.
The ALICE Collaboration acknowledges the following funding agencies for their support in building and running the ALICE detector:
A. I. Alikhanyan National Science Laboratory (Yerevan Physics Institute) Foundation (ANSL), State Committee of Science and World Federation of Scientists (WFS), Armenia;
Austrian Academy of Sciences, Austrian Science Fund (FWF): [M 2467-N36] and Nationalstiftung f\"{u}r Forschung, Technologie und Entwicklung, Austria;
Ministry of Communications and High Technologies, National Nuclear Research Center, Azerbaijan;
Conselho Nacional de Desenvolvimento Cient\'{\i}fico e Tecnol\'{o}gico (CNPq), Financiadora de Estudos e Projetos (Finep), Funda\c{c}\~{a}o de Amparo \`{a} Pesquisa do Estado de S\~{a}o Paulo (FAPESP) and Universidade Federal do Rio Grande do Sul (UFRGS), Brazil;
Bulgarian Ministry of Education and Science, within the National Roadmap for Research Infrastructures 2020-2027 (object CERN), Bulgaria;
Ministry of Education of China (MOEC) , Ministry of Science \& Technology of China (MSTC) and National Natural Science Foundation of China (NSFC), China;
Ministry of Science and Education and Croatian Science Foundation, Croatia;
Centro de Aplicaciones Tecnol\'{o}gicas y Desarrollo Nuclear (CEADEN), Cubaenerg\'{\i}a, Cuba;
Ministry of Education, Youth and Sports of the Czech Republic, Czech Republic;
The Danish Council for Independent Research | Natural Sciences, the VILLUM FONDEN and Danish National Research Foundation (DNRF), Denmark;
Helsinki Institute of Physics (HIP), Finland;
Commissariat \`{a} l'Energie Atomique (CEA) and Institut National de Physique Nucl\'{e}aire et de Physique des Particules (IN2P3) and Centre National de la Recherche Scientifique (CNRS), France;
Bundesministerium f\"{u}r Bildung und Forschung (BMBF) and GSI Helmholtzzentrum f\"{u}r Schwerionenforschung GmbH, Germany;
General Secretariat for Research and Technology, Ministry of Education, Research and Religions, Greece;
National Research, Development and Innovation Office, Hungary;
Department of Atomic Energy Government of India (DAE), Department of Science and Technology, Government of India (DST), University Grants Commission, Government of India (UGC) and Council of Scientific and Industrial Research (CSIR), India;
National Research and Innovation Agency - BRIN, Indonesia;
Istituto Nazionale di Fisica Nucleare (INFN), Italy;
Japanese Ministry of Education, Culture, Sports, Science and Technology (MEXT) and Japan Society for the Promotion of Science (JSPS) KAKENHI, Japan;
Consejo Nacional de Ciencia (CONACYT) y Tecnolog\'{i}a, through Fondo de Cooperaci\'{o}n Internacional en Ciencia y Tecnolog\'{i}a (FONCICYT) and Direcci\'{o}n General de Asuntos del Personal Academico (DGAPA), Mexico;
Nederlandse Organisatie voor Wetenschappelijk Onderzoek (NWO), Netherlands;
The Research Council of Norway, Norway;
Commission on Science and Technology for Sustainable Development in the South (COMSATS), Pakistan;
Pontificia Universidad Cat\'{o}lica del Per\'{u}, Peru;
Ministry of Education and Science, National Science Centre and WUT ID-UB, Poland;
Korea Institute of Science and Technology Information and National Research Foundation of Korea (NRF), Republic of Korea;
Ministry of Education and Scientific Research, Institute of Atomic Physics, Ministry of Research and Innovation and Institute of Atomic Physics and Universitatea Nationala de Stiinta si Tehnologie Politehnica Bucuresti, Romania;
Ministry of Education, Science, Research and Sport of the Slovak Republic, Slovakia;
National Research Foundation of South Africa, South Africa;
Swedish Research Council (VR) and Knut \& Alice Wallenberg Foundation (KAW), Sweden;
European Organization for Nuclear Research, Switzerland;
Suranaree University of Technology (SUT), National Science and Technology Development Agency (NSTDA) and National Science, Research and Innovation Fund (NSRF via PMU-B B05F650021), Thailand;
Turkish Energy, Nuclear and Mineral Research Agency (TENMAK), Turkey;
National Academy of  Sciences of Ukraine, Ukraine;
Science and Technology Facilities Council (STFC), United Kingdom;
National Science Foundation of the United States of America (NSF) and United States Department of Energy, Office of Nuclear Physics (DOE NP), United States of America.
In addition, individual groups or members have received support from:
Czech Science Foundation (grant no. 23-07499S), Czech Republic;
European Research Council, Strong 2020 - Horizon 2020 (grant nos. 950692, 824093), European Union;
ICSC - Centro Nazionale di Ricerca in High Performance Computing, Big Data and Quantum Computing, European Union - NextGenerationEU;
Academy of Finland (Center of Excellence in Quark Matter) (grant nos. 346327, 346328), Finland.

\end{acknowledgement}

\bibliographystyle{utphys}   
\bibliography{Kpi.bib}

\providecommand{\href}[2]{#2}\begingroup\raggedright\begin{thebibliography}{10}

\bibitem{Lisa:2005dd}
M.~A. Lisa, S.~Pratt, R.~Soltz, and U.~Wiedemann, ``{Femtoscopy in relativistic
  heavy ion collisions}'',
  \href{http://dx.doi.org/10.1146/annurev.nucl.55.090704.151533}{{\em Ann. Rev.
  Nucl. Part. Sci.} {\bfseries 55} (2005) 357--402},
\href{http://arxiv.org/abs/nucl-ex/0505014}{{\ttfamily arXiv:nucl-ex/0505014
  [nucl-ex]}}.

\bibitem{Abelev:2006gu}
{\bfseries STAR} Collaboration, B.~I. Abelev {\em et~al.}, ``{Neutral kaon
  interferometry in Au+Au collisions at $\sqrt{s_{\rm NN}} = 200$ GeV}'',
  \href{http://dx.doi.org/10.1103/PhysRevC.74.054902}{{\em Phys. Rev.}
  {\bfseries C74} (2006) 054902},
\href{http://arxiv.org/abs/nucl-ex/0608012}{{\ttfamily arXiv:nucl-ex/0608012
  [nucl-ex]}}.

\bibitem{PHENIX:2015jaj}
{\bfseries PHENIX} Collaboration, A.~Adare {\em et~al.}, ``{Systematic study of
  charged-pion and kaon femtoscopy in Au + Au collisions at
  $\sqrt{s_{_{NN}}}$=200 GeV}'',
  \href{http://dx.doi.org/10.1103/PhysRevC.92.034914}{{\em Phys. Rev. C}
  {\bfseries 92} (2015) 034914},
  \href{http://arxiv.org/abs/1504.05168}{{\ttfamily arXiv:1504.05168
  [nucl-ex]}}.

\bibitem{Abelev:2012ms}
{\bfseries ALICE} Collaboration, B.~Abelev {\em et~al.}, ``{$\rm K^0_s-\rm
  K^0_s$ correlations in pp collisions at $\sqrt{s}=7$ TeV from the LHC ALICE
  experiment}'', \href{http://dx.doi.org/10.1016/j.physletb.2012.09.013}{{\em
  Phys. Lett.} {\bfseries B717} (2012) 151--161},
\href{http://arxiv.org/abs/1206.2056}{{\ttfamily arXiv:1206.2056 [hep-ex]}}.

\bibitem{Abelev:2012sq}
{\bfseries ALICE} Collaboration, B.~Abelev {\em et~al.}, ``{Charged kaon
  femtoscopic correlations in pp collisions at $\sqrt{s}=7$ TeV}'',
  \href{http://dx.doi.org/10.1103/PhysRevD.87.052016}{{\em Phys. Rev.}
  {\bfseries D87} (2013) 052016},
\href{http://arxiv.org/abs/1212.5958}{{\ttfamily arXiv:1212.5958 [hep-ex]}}.

\bibitem{Adam:2015vja}
{\bfseries ALICE} Collaboration, J.~Adam {\em et~al.}, ``{One-dimensional pion,
  kaon, and proton femtoscopy in Pb--Pb collisions at $\sqrt{s_{\rm NN}}$ =2.76
  TeV}'', \href{http://dx.doi.org/10.1103/PhysRevC.92.054908}{{\em Phys. Rev.}
  {\bfseries C92} (2015) 054908},
\href{http://arxiv.org/abs/1506.07884}{{\ttfamily arXiv:1506.07884 [nucl-ex]}}.

\bibitem{ALICE:2021ovd}
{\bfseries ALICE} Collaboration, S.~Acharya {\em et~al.}, ``{K$^0_{\rm
  s}$K$^0_{\rm s}$ and K$^0_{\rm s}$K$^\pm$ femtoscopy in pp collisions at
  $\sqrt{s}$=5.02 and 13 TeV}'',
  \href{http://dx.doi.org/10.1016/j.physletb.2022.137335}{{\em Phys. Lett. B}
  {\bfseries 833} (2022) 137335},
  \href{http://arxiv.org/abs/2111.06611}{{\ttfamily arXiv:2111.06611
  [nucl-ex]}}.

\bibitem{Fabbietti:2020bfg}
L.~Fabbietti, V.~Mantovani~Sarti, and O.~Vazquez~Doce, ``{Study of the Strong
  Interaction Among Hadrons with Correlations at the LHC}'',
  \href{http://dx.doi.org/10.1146/annurev-nucl-102419-034438}{{\em Ann. Rev.
  Nucl. Part. Sci.} {\bfseries 71} (2021) 377--402},
  \href{http://arxiv.org/abs/2012.09806}{{\ttfamily arXiv:2012.09806
  [nucl-ex]}}.

\bibitem{Acharya:2018kpo}
{\bfseries ALICE} Collaboration, S.~Acharya {\em et~al.}, ``{Measuring
  K$^0_{\rm S}$K$^{\rm{\pm}}$ interactions using pp collisions at $\sqrt{s}=7$
  TeV}'', \href{http://dx.doi.org/10.1016/j.physletb.2018.12.033}{{\em Phys.
  Lett.} {\bfseries B790} (2019) 22--34},
\href{http://arxiv.org/abs/1809.07899}{{\ttfamily arXiv:1809.07899 [nucl-ex]}}.

\bibitem{Acharya:2017jks}
{\bfseries ALICE} Collaboration, S.~Acharya {\em et~al.}, ``{Measuring
  K$^0_{\rm S}$K$^{\rm \pm}$ interactions using Pb--Pb collisions at
  ${\sqrt{s_{\rm NN}}=2.76}$ TeV}'',
  \href{http://dx.doi.org/10.1016/j.physletb.2017.09.009}{{\em Phys. Lett. B}
  {\bfseries 774} (2017) 64--77},
  \href{http://arxiv.org/abs/1705.04929}{{\ttfamily arXiv:1705.04929
  [nucl-ex]}}.

\bibitem{Santopinto:2006my}
E.~Santopinto and G.~Galata, ``{Spectroscopy of tetraquark states}'',
  \href{http://dx.doi.org/10.1103/PhysRevC.75.045206}{{\em Phys. Rev.}
  {\bfseries C75} (2007) 045206},
\href{http://arxiv.org/abs/hep-ph/0605333}{{\ttfamily arXiv:hep-ph/0605333
  [hep-ph]}}.

\bibitem{Jaffe:1976ig}
R.~L. Jaffe, ``{Multi-Quark Hadrons. 1. The Phenomenology of ${\rm
  qq\overline{q}\overline{q}}$ Mesons}'',
\href{http://dx.doi.org/10.1103/PhysRevD.15.267}{{\em Phys. Rev.} {\bfseries
  D15} (1977) 267}.

\bibitem{Alford:2000mm}
M.~G. Alford and R.~L. Jaffe, ``{Insight into the scalar mesons from a lattice
  calculation}'', \href{http://dx.doi.org/10.1016/S0550-3213(00)00155-3}{{\em
  Nucl. Phys.} {\bfseries B578} (2000) 367--382},
\href{http://arxiv.org/abs/hep-lat/0001023}{{\ttfamily arXiv:hep-lat/0001023
  [hep-lat]}}.

\bibitem{Narison:2008nj}
S.~Narison, ``{Light scalar mesons in QCD}'',
  \href{http://dx.doi.org/10.1016/j.nuclphysbps.2008.12.069}{{\em Nucl. Phys. B
  Proc. Suppl.} {\bfseries 186} (2009) 306--311},
  \href{http://arxiv.org/abs/0811.0563}{{\ttfamily arXiv:0811.0563 [hep-ph]}}.

\bibitem{Achasov:2017zhy}
N.~Achasov and A.~Kiselev, ``{Light scalar mesons and two-kaon correlation
  functions}'', \href{http://dx.doi.org/10.1103/PhysRevD.97.036015}{{\em Phys.
  Rev. D} {\bfseries 97} (2018) 036015},
  \href{http://arxiv.org/abs/1711.08777}{{\ttfamily arXiv:1711.08777
  [hep-ph]}}.

\bibitem{Azizi:2019kzj}
K.~Azizi, B.~Barsbay, and H.~Sundu, ``{Light scalar $K_{0}^{*}(700)$ meson in
  vacuum and a hot medium}'',
  \href{http://dx.doi.org/10.1103/PhysRevD.100.094041}{{\em Phys. Rev. D}
  {\bfseries 100} (2019) 094041},
  \href{http://arxiv.org/abs/1909.00716}{{\ttfamily arXiv:1909.00716
  [hep-ph]}}.

\bibitem{Dudek:2016cru}
{\bfseries Hadron Spectrum} Collaboration, J.~J. Dudek, R.~G. Edwards, and
  D.~J. Wilson, ``{An $a_0$ resonance in strongly coupled $\pi \eta$,
  $K\overline{K}$ scattering from lattice QCD}'',
  \href{http://dx.doi.org/10.1103/PhysRevD.93.094506}{{\em Phys. Rev. D}
  {\bfseries 93} (2016) 094506},
  \href{http://arxiv.org/abs/1602.05122}{{\ttfamily arXiv:1602.05122
  [hep-ph]}}.

\bibitem{Briceno:2016mjc}
R.~A. Briceno, J.~J. Dudek, R.~G. Edwards, and D.~J. Wilson, ``{Isoscalar
  $\pi\pi$ scattering and the $\sigma$ meson resonance from QCD}'',
  \href{http://dx.doi.org/10.1103/PhysRevLett.118.022002}{{\em Phys. Rev.
  Lett.} {\bfseries 118} (2017) 022002},
  \href{http://arxiv.org/abs/1607.05900}{{\ttfamily arXiv:1607.05900
  [hep-ph]}}.

\bibitem{Guo:2013nja}
F.-K. Guo, L.~Liu, U.-G. Meissner, and P.~Wang, ``{Tetraquarks, hadronic
  molecules, meson-meson scattering and disconnected contributions in lattice
  QCD}'', \href{http://dx.doi.org/10.1103/PhysRevD.88.074506}{{\em Phys. Rev.
  D} {\bfseries 88} (2013) 074506},
  \href{http://arxiv.org/abs/1308.2545}{{\ttfamily arXiv:1308.2545 [hep-lat]}}.

\bibitem{Workman:2022ynf}
{\bfseries Particle Data Group} Collaboration, R.~L. Workman and Others,
  ``{Review of Particle Physics}'',
  \href{http://dx.doi.org/10.1093/ptep/ptac097}{{\em PTEP} {\bfseries 2022}
  (2022) 083C01}.

\bibitem{Humanic:2018brf}
T.~J. Humanic, ``{Feasibility of studying the K$^*_0(700)$ resonance using
  $\pi^{\rm \pm}$K$^0_{\rm S}$ femtoscopy}'',
  \href{http://dx.doi.org/10.1088/1361-6471/ab0b72}{{\em J. Phys. G} {\bfseries
  46} (2019) 055001}, \href{http://arxiv.org/abs/1810.10959}{{\ttfamily
  arXiv:1810.10959 [hep-ph]}}.

\bibitem{Aamodt:2008zz}
{\bfseries ALICE} Collaboration, K.~Aamodt {\em et~al.}, ``{The ALICE
  experiment at the CERN LHC}'',
\href{http://dx.doi.org/10.1088/1748-0221/3/08/S08002}{{\em JINST} {\bfseries
  3} (2008) S08002}.

\bibitem{ALICE:2022wpn}
{\bfseries ALICE} Collaboration, S.~Acharya {\em et~al.}, ``{The ALICE
  experiment: a journey through QCD}'',
  \href{http://dx.doi.org/10.1140/epjc/s10052-024-12935-y}{{\em Eur. Phys. J.
  C} {\bfseries 84} (2024) 813},
  \href{http://arxiv.org/abs/2211.04384}{{\ttfamily arXiv:2211.04384
  [nucl-ex]}}.

\bibitem{Alessandro:2006yt}
{\bfseries ALICE} Collaboration, B.~Alessandro {\em et~al.}, ``{ALICE: Physics
  performance report, volume II}'',
\href{http://dx.doi.org/10.1088/0954-3899/32/10/001}{{\em J. Phys.} {\bfseries
  G32} (2006) 1295--2040}.

\bibitem{Abelev:2013vea}
{\bfseries ALICE} Collaboration, B.~Abelev {\em et~al.}, ``{Centrality
  dependence of $\pi$, K, p production in Pb--Pb collisions at $\sqrt{s_{\rm
  NN}}$ = 2.76 TeV}'', \href{http://dx.doi.org/10.1103/PhysRevC.88.044910}{{\em
  Phys. Rev.} {\bfseries C88} (2013) 044910},
\href{http://arxiv.org/abs/1303.0737}{{\ttfamily arXiv:1303.0737 [hep-ex]}}.

\bibitem{Abelev:2013qoq}
{\bfseries ALICE} Collaboration, B.~Abelev {\em et~al.}, ``{Centrality
  determination of Pb--Pb collisions at $\sqrt{s_{\rm NN}}$=2.76~TeV with
  ALICE}'', \href{http://dx.doi.org/10.1103/PhysRevC.88.044909}{{\em Phys.
  Rev.} {\bfseries C88} (2013) 044909},
\href{http://arxiv.org/abs/1301.4361}{{\ttfamily arXiv:1301.4361 [nucl-ex]}}.

\bibitem{ALICE:2020swj}
{\bfseries ALICE} Collaboration, S.~Acharya {\em et~al.}, ``{Pseudorapidity
  distributions of charged particles as a function of mid- and forward rapidity
  multiplicities in pp collisions at $\sqrt{s}$~=~5.02, 7 and 13 TeV}'',
  \href{http://dx.doi.org/10.1140/epjc/s10052-021-09349-5}{{\em Eur. Phys. J.
  C} {\bfseries 81} (2021) 630},
  \href{http://arxiv.org/abs/2009.09434}{{\ttfamily arXiv:2009.09434
  [nucl-ex]}}.

\bibitem{Alme:2010ke}
J.~Alme {\em et~al.}, ``{The ALICE TPC, a large 3-dimensional tracking device
  with fast readout for ultra-high multiplicity events}'',
  \href{http://dx.doi.org/10.1016/j.nima.2010.04.042}{{\em Nucl. Instrum.
  Meth.} {\bfseries A622} (2010) 316--367},
\href{http://arxiv.org/abs/1001.1950}{{\ttfamily arXiv:1001.1950
  [physics.ins-det]}}.

\bibitem{Abelev:2014ffa}
{\bfseries ALICE} Collaboration, B.~B. Abelev {\em et~al.}, ``{Performance of
  the ALICE Experiment at the CERN LHC}'',
  \href{http://dx.doi.org/10.1142/S0217751X14300440}{{\em Int. J. Mod. Phys.}
  {\bfseries A29} (2014) 1430044},
\href{http://arxiv.org/abs/1402.4476}{{\ttfamily arXiv:1402.4476 [nucl-ex]}}.

\bibitem{Acharya:2019idg}
{\bfseries ALICE} Collaboration, S.~Acharya {\em et~al.}, ``{Event-shape and
  multiplicity dependence of freeze-out radii in pp collisions at $ \sqrt{s} $
  = 7 TeV}'', \href{http://dx.doi.org/10.1007/JHEP09(2019)108}{{\em JHEP}
  {\bfseries 09} (2019) 108}, \href{http://arxiv.org/abs/1901.05518}{{\ttfamily
  arXiv:1901.05518 [nucl-ex]}}.

\bibitem{Akindinov:2013tea}
A.~Akindinov {\em et~al.}, ``{Performance of the ALICE Time-Of-Flight detector
  at the LHC}'',
\href{http://dx.doi.org/10.1140/epjp/i2013-13044-x}{{\em Eur. Phys. J. Plus}
  {\bfseries 128} (2013) 44}.

\bibitem{Aamodt:2011zz}
{\bfseries ALICE} Collaboration, K.~Aamodt, ``{$\pi^0$ and $\eta$
  reconstruction from photon conversions in ALICE for first pp collisions at
  the LHC}'', \href{http://dx.doi.org/10.1088/1742-6596/270/1/012035}{{\em J.
  Phys. Conf. Ser.} {\bfseries 270} (2011) 012035}.

\bibitem{Sjostrand:2006za}
T.~Sjostrand, S.~Mrenna, and P.~Skands, ``{PYTHIA 6.4 Physics and Manual}'',
  {\em JHEP} {\bfseries 05} (2006) 026,
\href{http://arxiv.org/abs/hep-ph/0603175}{{\ttfamily arXiv:hep-ph/0603175}}.

\bibitem{Skands:2014pea}
P.~Skands, S.~Carrazza, and J.~Rojo, ``{Tuning PYTHIA 8.1: the Monash 2013
  Tune}'', \href{http://dx.doi.org/10.1140/epjc/s10052-014-3024-y}{{\em Eur.
  Phys. J. C} {\bfseries 74} (2014) 3024},
  \href{http://arxiv.org/abs/1404.5630}{{\ttfamily arXiv:1404.5630 [hep-ph]}}.

\bibitem{Brun:1994aa}
R.~Brun, F.~Bruyant, F.~Carminati, S.~Giani, M.~Maire, A.~McPherson,
  G.~Patrick, and L.~Urban, ``{GEANT Detector Description and Simulation
  Tool}'',
\href{http://dx.doi.org/1}{{\em CERN-W5013} {\bfseries 1} (1994) 1}.

\bibitem{ALICE:2011kmy}
{\bfseries ALICE} Collaboration, K.~Aamodt {\em et~al.}, ``{Femtoscopy of pp
  collisions at $\sqrt{s}=0.9$ and 7 TeV at the LHC with two-pion Bose-Einstein
  correlations}'', \href{http://dx.doi.org/10.1103/PhysRevD.84.112004}{{\em
  Phys. Rev. D} {\bfseries 84} (2011) 112004},
  \href{http://arxiv.org/abs/1101.3665}{{\ttfamily arXiv:1101.3665 [hep-ex]}}.

\bibitem{ALICE:2019gcn}
{\bfseries ALICE} Collaboration, S.~Acharya {\em et~al.}, ``{Scattering studies
  with low-energy kaon-proton femtoscopy in proton-proton collisions at the
  LHC}'', \href{http://dx.doi.org/10.1103/PhysRevLett.124.092301}{{\em Phys.
  Rev. Lett.} {\bfseries 124} (2020) 092301},
  \href{http://arxiv.org/abs/1905.13470}{{\ttfamily arXiv:1905.13470
  [nucl-ex]}}.

\bibitem{ALICE:2021cpv}
{\bfseries ALICE} Collaboration, S.~Acharya {\em et~al.}, ``{Experimental
  Evidence for an Attractive p-$\phi$ Interaction}'',
  \href{http://dx.doi.org/10.1103/PhysRevLett.127.172301}{{\em Phys. Rev.
  Lett.} {\bfseries 127} (2021) 172301},
  \href{http://arxiv.org/abs/2105.05578}{{\ttfamily arXiv:2105.05578
  [nucl-ex]}}.

\bibitem{Lednicky:1981su}
R.~Lednicky and V.~Lyuboshits, ``{Final State Interaction Effect on Pairing
  Correlations Between Particles with Small Relative Momenta}'',
{\em Sov. J. Nucl. Phys.} {\bfseries 35} (1982) 770.

\bibitem{Lednicky:2005af}
R.~Lednicky, ``{Correlation femtoscopy}'',
  \href{http://dx.doi.org/10.1016/j.nuclphysa.2006.06.040}{{\em Nucl. Phys.}
  {\bfseries A774} (2006) 189--198},
\href{http://arxiv.org/abs/nucl-th/0510020}{{\ttfamily arXiv:nucl-th/0510020
  [nucl-th]}}.

\bibitem{BES:2010soq}
{\bfseries BES} Collaboration, M.~Ablikim {\em et~al.}, ``{Observation of
  charged $\kappa$ in $J/\psi\rightarrow K^*(892)^{\pm}K_s \pi^{\pm}$,
  $K^*(892)^{\pm} \rightarrow K_s\pi^{\pm}$ at BESII}'',
  \href{http://dx.doi.org/10.1016/j.physletb.2011.03.011}{{\em Phys. Lett. B}
  {\bfseries 698} (2011) 183--190},
  \href{http://arxiv.org/abs/1008.4489}{{\ttfamily arXiv:1008.4489 [hep-ex]}}.

\bibitem{E791:2002xlc}
{\bfseries E791} Collaboration, E.~M. Aitala {\em et~al.}, ``{Dalitz Plot
  Analysis of the Decay $D^+ \to K^- \pi^+ \pi^+$ and the Study of the $K\pi$
  Scalar Amplitudes}'',
  \href{http://dx.doi.org/10.1103/PhysRevLett.89.121801}{{\em Phys. Rev. Lett.}
  {\bfseries 89} (2002) 121801},
  \href{http://arxiv.org/abs/hep-ex/0204018}{{\ttfamily arXiv:hep-ex/0204018}}.

\bibitem{james1971}
W.~T. Eadie {\em et~al.}, {\em Statistical Methods in Experimental Physics}.
\newblock North Holland, Amsterdam, 1971.

\bibitem{Bohm:2004zi}
A.~R. Bohm and Y.~Sato, ``{Relativistic resonances: Their masses, widths,
  lifetimes, superposition, and causal evolution}'',
  \href{http://dx.doi.org/10.1103/PhysRevD.71.085018}{{\em Phys. Rev. D}
  {\bfseries 71} (2005) 085018},
  \href{http://arxiv.org/abs/hep-ph/0412106}{{\ttfamily arXiv:hep-ph/0412106}}.

\bibitem{Humanic:2008nt}
T.~J. Humanic, ``{Hadronic observables from Au+Au collisions at $\sqrt{s_{\rm
  NN}}$ =200 GeV and Pb+Pb collisions at $\sqrt{s_{\rm NN}}$ =5.5 TeV from a
  simple kinematic model}'',
  \href{http://dx.doi.org/10.1103/PhysRevC.79.044902}{{\em Phys. Rev.}
  {\bfseries C79} (2009) 044902},
\href{http://arxiv.org/abs/0810.0621}{{\ttfamily arXiv:0810.0621 [nucl-th]}}.

\end{thebibliography}\endgroup

\newpage
\appendix

%
%

\section{The ALICE Collaboration}
\label{app:collab}
\begin{flushleft} 
\small

S.~Acharya\,\orcidlink{0000-0002-9213-5329}\,$^{\rm 128}$, 
D.~Adamov\'{a}\,\orcidlink{0000-0002-0504-7428}\,$^{\rm 87}$, 
G.~Aglieri Rinella\,\orcidlink{0000-0002-9611-3696}\,$^{\rm 33}$, 
L.~Aglietta$^{\rm 25}$, 
M.~Agnello\,\orcidlink{0000-0002-0760-5075}\,$^{\rm 30}$, 
N.~Agrawal\,\orcidlink{0000-0003-0348-9836}\,$^{\rm 26}$, 
Z.~Ahammed\,\orcidlink{0000-0001-5241-7412}\,$^{\rm 136}$, 
S.~Ahmad\,\orcidlink{0000-0003-0497-5705}\,$^{\rm 16}$, 
S.U.~Ahn\,\orcidlink{0000-0001-8847-489X}\,$^{\rm 72}$, 
I.~Ahuja\,\orcidlink{0000-0002-4417-1392}\,$^{\rm 38}$
% ...

\section*{Affiliation Notes}

$^{\rm I}$ Deceased\\
$^{\rm II}$ Also at: Max-Planck-Institut fur Physik, Munich, Germany\\
$^{\rm III}$ Also at: Italian National Agency for New Technologies, Energy and Sustainable Economic Development (ENEA), Bologna, Italy\\
$^{\rm IV}$ Also at: Dipartimento DET del Politecnico di Torino, Turin, Italy\\
$^{\rm V}$ Also at: Yildiz Technical University, Istanbul, T\"{u}rkiye\\
$^{\rm VI}$ Also at: Department of Applied Physics, Aligarh Muslim University, Aligarh, India\\
$^{\rm VII}$ Also at: Institute of Theoretical Physics, University of Wroclaw, Poland\\
$^{\rm VIII}$ Also at: An institution covered by a cooperation agreement with CERN\\

\section*{Collaboration Institutes}

$^{1}$ A.I. Alikhanyan National Science Laboratory (Yerevan Physics Institute) Foundation, Yerevan, Armenia\\
$^{2}$ AGH University of Krakow, Cracow, Poland\\
$^{3}$ Bogolyubov Institute for Theoretical Physics, National Academy of Sciences of Ukraine, Kiev, Ukraine\\
$^{4}$ Bose Institute, Department of Physics  and Centre for Astroparticle Physics and Space Science (CAPSS), Kolkata, India\\
$^{5}$ California Polytechnic State University, San Luis Obispo, California, United States\\
$^{6}$ Central China Normal University, Wuhan, China\\
$^{7}$ Centro de Aplicaciones Tecnol\'{o}gicas y Desarrollo Nuclear (CEADEN), Havana, Cuba\\
$^{8}$ Centro de Investigaci\'{o}n y de Estudios Avanzados (CINVESTAV), Mexico City and M\'{e}rida, Mexico\\
$^{9}$ Chicago State University, Chicago, Illinois, United States\\
$^{10}$ China Institute of Atomic Energy, Beijing, China\\
$^{11}$ China University of Geosciences, Wuhan, China\\
$^{12}$ Chungbuk National University, Cheongju, Republic of Korea\\
$^{13}$ Comenius University Bratislava, Faculty of Mathematics, Physics and Informatics, Bratislava, Slovak Republic\\
$^{14}$ COMSATS University Islamabad, Islamabad, Pakistan\\
$^{15}$ Creighton University, Omaha, Nebraska, United States\\
$^{16}$ Department of Physics, Aligarh Muslim University, Aligarh, India\\
$^{17}$ Department of Physics, Pusan National University, Pusan, Republic of Korea\\
$^{18}$ Department of Physics, Sejong University, Seoul, Republic of Korea\\
$^{19}$ Department of Physics, University of California, Berkeley, California, United States\\
$^{20}$ Department of Physics, University of Oslo, Oslo, Norway\\
$^{21}$ Department of Physics and Technology, University of Bergen, Bergen, Norway\\
$^{22}$ Dipartimento di Fisica, Universit\`{a} di Pavia, Pavia, Italy\\
$^{23}$ Dipartimento di Fisica dell'Universit\`{a} and Sezione INFN, Cagliari, Italy\\
$^{24}$ Dipartimento di Fisica dell'Universit\`{a} and Sezione INFN, Trieste, Italy\\
$^{25}$ Dipartimento di Fisica dell'Universit\`{a} and Sezione INFN, Turin, Italy\\
$^{26}$ Dipartimento di Fisica e Astronomia dell'Universit\`{a} and Sezione INFN, Bologna, Italy\\
$^{27}$ Dipartimento di Fisica e Astronomia dell'Universit\`{a} and Sezione INFN, Catania, Italy\\
$^{28}$ Dipartimento di Fisica e Astronomia dell'Universit\`{a} and Sezione INFN, Padova, Italy\\
$^{29}$ Dipartimento di Fisica `E.R.~Caianiello' dell'Universit\`{a} and Gruppo Collegato INFN, Salerno, Italy\\
$^{30}$ Dipartimento DISAT del Politecnico and Sezione INFN, Turin, Italy\\
$^{31}$ Dipartimento di Scienze MIFT, Universit\`{a} di Messina, Messina, Italy\\
$^{32}$ Dipartimento Interateneo di Fisica `M.~Merlin' and Sezione INFN, Bari, Italy\\
$^{33}$ European Organization for Nuclear Research (CERN), Geneva, Switzerland\\
$^{34}$ Faculty of Electrical Engineering, Mechanical Engineering and Naval Architecture, University of Split, Split, Croatia\\
$^{35}$ Faculty of Engineering and Science, Western Norway University of Applied Sciences, Bergen, Norway\\
$^{36}$ Faculty of Nuclear Sciences and Physical Engineering, Czech Technical University in Prague, Prague, Czech Republic\\
$^{37}$ Faculty of Physics, Sofia University, Sofia, Bulgaria\\
$^{38}$ Faculty of Science, P.J.~\v{S}af\'{a}rik University, Ko\v{s}ice, Slovak Republic\\
$^{39}$ Frankfurt Institute for Advanced Studies, Johann Wolfgang Goethe-Universit\"{a}t Frankfurt, Frankfurt, Germany\\
$^{40}$ Fudan University, Shanghai, China\\
$^{41}$ Gangneung-Wonju National University, Gangneung, Republic of Korea\\
$^{42}$ Gauhati University, Department of Physics, Guwahati, India\\
$^{43}$ Helmholtz-Institut f\"{u}r Strahlen- und Kernphysik, Rheinische Friedrich-Wilhelms-Universit\"{a}t Bonn, Bonn, Germany\\
$^{44}$ Helsinki Institute of Physics (HIP), Helsinki, Finland\\
$^{45}$ High Energy Physics Group,  Universidad Aut\'{o}noma de Puebla, Puebla, Mexico\\
$^{46}$ Horia Hulubei National Institute of Physics and Nuclear Engineering, Bucharest, Romania\\
$^{47}$ HUN-REN Wigner Research Centre for Physics, Budapest, Hungary\\
$^{48}$ Indian Institute of Technology Bombay (IIT), Mumbai, India\\
$^{49}$ Indian Institute of Technology Indore, Indore, India\\
$^{50}$ INFN, Laboratori Nazionali di Frascati, Frascati, Italy\\
$^{51}$ INFN, Sezione di Bari, Bari, Italy\\
$^{52}$ INFN, Sezione di Bologna, Bologna, Italy\\
$^{53}$ INFN, Sezione di Cagliari, Cagliari, Italy\\
$^{54}$ INFN, Sezione di Catania, Catania, Italy\\
$^{55}$ INFN, Sezione di Padova, Padova, Italy\\
$^{56}$ INFN, Sezione di Pavia, Pavia, Italy\\
$^{57}$ INFN, Sezione di Torino, Turin, Italy\\
$^{58}$ INFN, Sezione di Trieste, Trieste, Italy\\
$^{59}$ Inha University, Incheon, Republic of Korea\\
$^{60}$ Institute for Gravitational and Subatomic Physics (GRASP), Utrecht University/Nikhef, Utrecht, Netherlands\\
$^{61}$ Institute of Experimental Physics, Slovak Academy of Sciences, Ko\v{s}ice, Slovak Republic\\
$^{62}$ Institute of Physics, Homi Bhabha National Institute, Bhubaneswar, India\\
$^{63}$ Institute of Physics of the Czech Academy of Sciences, Prague, Czech Republic\\
$^{64}$ Institute of Space Science (ISS), Bucharest, Romania\\
$^{65}$ Institut f\"{u}r Kernphysik, Johann Wolfgang Goethe-Universit\"{a}t Frankfurt, Frankfurt, Germany\\
$^{66}$ Instituto de Ciencias Nucleares, Universidad Nacional Aut\'{o}noma de M\'{e}xico, Mexico City, Mexico\\
$^{67}$ Instituto de F\'{i}sica, Universidade Federal do Rio Grande do Sul (UFRGS), Porto Alegre, Brazil\\
$^{68}$ Instituto de F\'{\i}sica, Universidad Nacional Aut\'{o}noma de M\'{e}xico, Mexico City, Mexico\\
$^{69}$ iThemba LABS, National Research Foundation, Somerset West, South Africa\\
$^{70}$ Jeonbuk National University, Jeonju, Republic of Korea\\
$^{71}$ Johann-Wolfgang-Goethe Universit\"{a}t Frankfurt Institut f\"{u}r Informatik, Fachbereich Informatik und Mathematik, Frankfurt, Germany\\
$^{72}$ Korea Institute of Science and Technology Information, Daejeon, Republic of Korea\\
$^{73}$ KTO Karatay University, Konya, Turkey\\
$^{74}$ Laboratoire de Physique Subatomique et de Cosmologie, Universit\'{e} Grenoble-Alpes, CNRS-IN2P3, Grenoble, France\\
$^{75}$ Lawrence Berkeley National Laboratory, Berkeley, California, United States\\
$^{76}$ Lund University Department of Physics, Division of Particle Physics, Lund, Sweden\\
$^{77}$ Nagasaki Institute of Applied Science, Nagasaki, Japan\\
$^{78}$ Nara Women{'}s University (NWU), Nara, Japan\\
$^{79}$ National and Kapodistrian University of Athens, School of Science, Department of Physics , Athens, Greece\\
$^{80}$ National Centre for Nuclear Research, Warsaw, Poland\\
$^{81}$ National Institute of Science Education and Research, Homi Bhabha National Institute, Jatni, India\\
$^{82}$ National Nuclear Research Center, Baku, Azerbaijan\\
$^{83}$ National Research and Innovation Agency - BRIN, Jakarta, Indonesia\\
$^{84}$ Niels Bohr Institute, University of Copenhagen, Copenhagen, Denmark\\
$^{85}$ Nikhef, National institute for subatomic physics, Amsterdam, Netherlands\\
$^{86}$ Nuclear Physics Group, STFC Daresbury Laboratory, Daresbury, United Kingdom\\
$^{87}$ Nuclear Physics Institute of the Czech Academy of Sciences, Husinec-\v{R}e\v{z}, Czech Republic\\
$^{88}$ Oak Ridge National Laboratory, Oak Ridge, Tennessee, United States\\
$^{89}$ Ohio State University, Columbus, Ohio, United States\\
$^{90}$ Physics department, Faculty of science, University of Zagreb, Zagreb, Croatia\\
$^{91}$ Physics Department, Panjab University, Chandigarh, India\\
$^{92}$ Physics Department, University of Jammu, Jammu, India\\
$^{93}$ Physics Program and International Institute for Sustainability with Knotted Chiral Meta Matter (SKCM2), Hiroshima University, Hiroshima, Japan\\
$^{94}$ Physikalisches Institut, Eberhard-Karls-Universit\"{a}t T\"{u}bingen, T\"{u}bingen, Germany\\
$^{95}$ Physikalisches Institut, Ruprecht-Karls-Universit\"{a}t Heidelberg, Heidelberg, Germany\\
$^{96}$ Physik Department, Technische Universit\"{a}t M\"{u}nchen, Munich, Germany\\
$^{97}$ Politecnico di Bari and Sezione INFN, Bari, Italy\\
$^{98}$ Research Division and ExtreMe Matter Institute EMMI, GSI Helmholtzzentrum f\"ur Schwerionenforschung GmbH, Darmstadt, Germany\\
$^{99}$ Saga University, Saga, Japan\\
$^{100}$ Saha Institute of Nuclear Physics, Homi Bhabha National Institute, Kolkata, India\\
$^{101}$ School of Physics and Astronomy, University of Birmingham, Birmingham, United Kingdom\\
$^{102}$ Secci\'{o}n F\'{\i}sica, Departamento de Ciencias, Pontificia Universidad Cat\'{o}lica del Per\'{u}, Lima, Peru\\
$^{103}$ Stefan Meyer Institut f\"{u}r Subatomare Physik (SMI), Vienna, Austria\\
$^{104}$ SUBATECH, IMT Atlantique, Nantes Universit\'{e}, CNRS-IN2P3, Nantes, France\\
$^{105}$ Sungkyunkwan University, Suwon City, Republic of Korea\\
$^{106}$ Suranaree University of Technology, Nakhon Ratchasima, Thailand\\
$^{107}$ Technical University of Ko\v{s}ice, Ko\v{s}ice, Slovak Republic\\
$^{108}$ The Henryk Niewodniczanski Institute of Nuclear Physics, Polish Academy of Sciences, Cracow, Poland\\
$^{109}$ The University of Texas at Austin, Austin, Texas, United States\\
$^{110}$ Universidad Aut\'{o}noma de Sinaloa, Culiac\'{a}n, Mexico\\
$^{111}$ Universidade de S\~{a}o Paulo (USP), S\~{a}o Paulo, Brazil\\
$^{112}$ Universidade Estadual de Campinas (UNICAMP), Campinas, Brazil\\
$^{113}$ Universidade Federal do ABC, Santo Andre, Brazil\\
$^{114}$ Universitatea Nationala de Stiinta si Tehnologie Politehnica Bucuresti, Bucharest, Romania\\
$^{115}$ University of Cape Town, Cape Town, South Africa\\
$^{116}$ University of Derby, Derby, United Kingdom\\
$^{117}$ University of Houston, Houston, Texas, United States\\
$^{118}$ University of Jyv\"{a}skyl\"{a}, Jyv\"{a}skyl\"{a}, Finland\\
$^{119}$ University of Kansas, Lawrence, Kansas, United States\\
$^{120}$ University of Liverpool, Liverpool, United Kingdom\\
$^{121}$ University of Science and Technology of China, Hefei, China\\
$^{122}$ University of South-Eastern Norway, Kongsberg, Norway\\
$^{123}$ University of Tennessee, Knoxville, Tennessee, United States\\
$^{124}$ University of the Witwatersrand, Johannesburg, South Africa\\
$^{125}$ University of Tokyo, Tokyo, Japan\\
$^{126}$ University of Tsukuba, Tsukuba, Japan\\
$^{127}$ Universit\"{a}t M\"{u}nster, Institut f\"{u}r Kernphysik, M\"{u}nster, Germany\\
$^{128}$ Universit\'{e} Clermont Auvergne, CNRS/IN2P3, LPC, Clermont-Ferrand, France\\
$^{129}$ Universit\'{e} de Lyon, CNRS/IN2P3, Institut de Physique des 2 Infinis de Lyon, Lyon, France\\
$^{130}$ Universit\'{e} de Strasbourg, CNRS, IPHC UMR 7178, F-67000 Strasbourg, France, Strasbourg, France\\
$^{131}$ Universit\'{e} Paris-Saclay, Centre d'Etudes de Saclay (CEA), IRFU, D\'{e}partment de Physique Nucl\'{e}aire (DPhN), Saclay, France\\
$^{132}$ Universit\'{e}  Paris-Saclay, CNRS/IN2P3, IJCLab, Orsay, France\\
$^{133}$ Universit\`{a} degli Studi di Foggia, Foggia, Italy\\
$^{134}$ Universit\`{a} del Piemonte Orientale, Vercelli, Italy\\
$^{135}$ Universit\`{a} di Brescia, Brescia, Italy\\
$^{136}$ Variable Energy Cyclotron Centre, Homi Bhabha National Institute, Kolkata, India\\
$^{137}$ Warsaw University of Technology, Warsaw, Poland\\
$^{138}$ Wayne State University, Detroit, Michigan, United States\\
$^{139}$ Yale University, New Haven, Connecticut, United States\\
$^{140}$ Yonsei University, Seoul, Republic of Korea\\
$^{141}$  Zentrum  f\"{u}r Technologie und Transfer (ZTT), Worms, Germany\\
$^{142}$ Affiliated with an institute covered by a cooperation agreement with CERN\\
$^{143}$ Affiliated with an international laboratory covered by a cooperation agreement with CERN.\\

\end{flushleft}

\end{document}